\documentclass[aps,pre,preprint,showpacs,showkeys,floatfix]{revtex4-1}
\usepackage{graphicx}% Include figure files
\usepackage{dcolumn}% Align table columns on decimal point
\usepackage{bm}% bold math
\usepackage{float}
\usepackage{color}
\usepackage{amssymb}
\usepackage{amsmath}
\usepackage{color}
\usepackage{subcaption}
\usepackage{xcolor}
\usepackage[percent]{overpic}
\usepackage{enumitem}
\usepackage{natbib}
\usepackage{overpic}
\newcommand{\valmed}[1]{\langle #1 \rangle}

\usepackage{tikz}
\usetikzlibrary{shapes}

\definecolor{verdecito}{RGB}{0,168,89}
\definecolor{azulito}{RGB}{0,0,255}

\newcommand{\bluesolid}{\raisebox{2pt}{\tikz{\draw[blue,solid,line
        width=1pt](0,0) -- (5mm,0);}}}

\newcommand{\blacksolid}{\raisebox{2.5pt}{\tikz{\draw[black,solid,line
        width=1pt](0,0) -- (5mm,0);}}}

\newcommand{\bluedash}{\raisebox{2pt}{\tikz{\draw[blue,dashed,line
        width=1pt](0,0) -- (5mm,0);}}}

\newcommand{\reddash}{\raisebox{2pt}{\tikz{\draw[red,dashed,line
        width=1pt](0,0) -- (5mm,0);}}}

\newcommand{\bluedashdotted}{\raisebox{2pt}{\tikz{\draw[blue,dashdotted,line
        width=1pt](0,0) -- (5mm,0);}}}

\newcommand{\bluedotted}{\raisebox{2pt}{\tikz{\draw[blue,dotted,line
        width=1pt](0,0) -- (5mm,0);}}}

\newcommand{\bluesquare}{\raisebox{1.3pt}{\tikz{\node[draw,scale=0.41,regular
        polygon, regular polygon sides=4,blue](){};}}}

\newcommand{\redcircle}{\raisebox{0.5pt}{\tikz{\node[draw,scale=0.5,circle,red,line
        width=0.6pt](){};}}}

\begin{document}

\title{Controlling distant contacts to reduce disease spreading on disordered complex networks}

\author{Ignacio A. Perez} \email{ignacioperez@mdp.edu.ar}
\affiliation{Instituto de Investigaciones F\'isicas de Mar del Plata
  (IFIMAR)-Departamento de F\'isica, FCEyN, Universidad Nacional de
  Mar del Plata-CONICET, Funes 3350, (7600) Mar del Plata, Argentina.}

\author{Paul A. Trunfio} \affiliation{Physics Department and Center
  for Polymer Studies, Boston University, Boston, Massachusetts 02215,
  USA}

\author{Cristian E. La Rocca} \affiliation{Instituto de
  Investigaciones F\'isicas de Mar del Plata (IFIMAR)-Departamento de
  F\'isica, FCEyN, Universidad Nacional de Mar del Plata-CONICET,
  Funes 3350, (7600) Mar del Plata, Argentina.}  \affiliation{Physics
  Department and Center for Polymer Studies, Boston University,
  Boston, Massachusetts 02215, USA}

\author{Lidia A. Braunstein} \affiliation{Instituto de Investigaciones
  F\'isicas de Mar del Plata (IFIMAR)-Departamento de F\'isica, FCEyN,
  Universidad Nacional de Mar del Plata-CONICET, Funes 3350, (7600)
  Mar del Plata, Argentina.} \affiliation{Physics Department and Center
  for Polymer Studies, Boston University, Boston, Massachusetts 02215,
  USA}

\begin{abstract} 
  
\noindent
In real social networks, person-to-person interactions are known to be
heterogeneous, which can affect the way a disease spreads through a
population, reaches a tipping point in the fraction of infected
individuals, and becomes an epidemic. This property, called
\emph{disorder}, is usually associated with contact times between
individuals and can be modeled by a weighted network, where the
weights are related to normalized contact times $\omega$. In this
paper, we study the SIR model for disease spreading when both close and
distant types of interactions are present. We develop a
mitigation strategy that reduces only the time duration of distant
contacts, which are easier to alter in practice. Using branching
theory, supported by simulations, we found that the effectiveness of
the strategy increases when the density $f_1$ of close contacts
decreases. Moreover, we found a threshold $\tilde{f}_1 = T_c / \beta$
below which the strategy can bring the system from an epidemic to a
non-epidemic phase, even when close contacts have the longest time
durations.

\end{abstract}


\maketitle

\section{Introduction}
\label{intro}

\noindent

In recent centuries, changes in social contact patterns have caused
infectious diseases to propagate more rapidly and become more
widespread \cite{Ander-92}. The population growth in urban zones and
the increasing speed and efficiency of air travel have allowed
diseases to spread over long distances within months or even
weeks. Examples include the 1918 Spanish flu \cite{Joh-02}, the 2009
A(H1N1) influenza epidemic \cite{Fras-09}, the 2014 Ebola epidemic
\cite{Mer-15}, and the recent measles outbreak in Israel that
propagated to New York \cite{NBC-19}. Prior research has indicated that
due to the increased resistance of bacteria to drugs \cite{Vand-00},
climate change \cite{Mal-04,Mcm-06} and the deforestation of sylvan
areas \cite{Gre-07}, the number of diseases will continue to
increase. In this context, mathematical models allow epidemiologists
and sanitary authorities to understand propagation processes, predict
their effect on healthy populations, and evaluate the effectiveness of
different mitigation strategies. Although many models consider full
mixing, in which all individuals in a population interact with each
other \cite{Ander-92}, this assumption does not reflect a realistic
situation where an individual has a limited number of interactions and
where it can vary between individuals. For that reason, in recent
decades, researchers have begun to model epidemic processes using
complex networks, in which a node (an individual) has a probability
$P(k)$ of being connected with $k$ different nodes (neighbors) with
$k_{\rm min} \leq k \leq k_{\rm max}$, where $k_{\rm min}$ and $k_{\rm
  max}$ are the minimum and maximum connectivity,
respectively; they have found that connection patterns strongly affect
the spreading of a disease
\cite{New-02,Bocc-06,New-10,Cast-10,Pastor-15,Brau-17}.

The Susceptible-Infected-Recovered (SIR) model
\cite{Gra-83,Ander-92,New-02,Bocc-06} is a simple representation of
non-recurrent diseases, where individuals acquire permanent immunity
once they stop being ill. Examples include influenza A(N1H1), measles
and pertussis. In this model, an individual is either \emph{susceptible}
(able to be infected), \emph{infected} (can propagate the disease), or
\emph{recovered} (has either acquired an immunity or has died, thereby
no longer propagating the disease). In the discrete-time Reed-Frost
model \cite{Bai-75}, at each time step, infected individuals spread the
disease to susceptible neighbors, with probability $\beta$, and recover
$t_r$ time steps after being infected. The effective probability of
infection is, thus, given by the transmissibility $T = 1 - (1 -
\beta)^{t_r}$. The process ends when there are no more infected
individuals; the system has reached the steady state. The SIR model, at
the steady state, exhibits a second-order phase transition where the
fraction $R$ of recovered individuals is the order parameter, while $T$
is the control parameter. Below a critical threshold $T = T_c$, the
disease reaches only a small fraction of the population, but when $T >
T_c$ it becomes an epidemic \cite{New-02,Mil-07,Ken-07,Lag-09}. Studies
have shown that the steady state of the SIR model is related to link
percolation \cite{Gra-83,New-02,Ken-07,Mey-07}. In the SIR model, links
are occupied with probability $p$ since the propagation of the disease
from an infected to a susceptible individual, is equivalent to occupying
that link via link percolation (i.e., $T \equiv p$). Above a critical
threshold $p = p_c$, a giant component (GC) of the same order of
magnitude than the system size $N$ emerges, whereas below $p_c$ there
are only finite clusters. The fraction $P_{\infty}$ of nodes belonging
to the GC is the order parameter of a second-order phase transition,
with $R$ from the SIR model mapping into $P_{\infty}$
\cite{New-02}. Because the SIR process only produces one cluster of
nodes (those reached by the disease), realizations with $R < s_c$ are
neglected for the mapping to exist~\cite{Lag-09}. For complex networks
$p_c = 1 /(\kappa - 1) = T_c$, where $\kappa = \valmed{k^2} /
\valmed{k}$; $\valmed{k}$ and $\valmed{k^2}$ are the first and second
moments of the distribution $P(k)$, respectively
\cite{Gra-83,New-02,Ken-07}.

There are different strategies proposed to contain the spreading
of diseases. Vaccination is one of the more studied and it is highly
efficient in providing immunity \cite{Fer-06,Ban-06,DiMu-18}, although
vaccines are often expensive and not always available. In this
context, non-pharmacological strategies are needed to protect
populations. For instance, quarantine is one of the most effective,
however complete isolation has a deleterious effect on the economy of a
region, and it is difficult to implement in a large population. Thus,
``social distancing'' \cite{Gro-06,Lag-11,Buo-12,Val-12,Buo-13}, i.e.,
reducing the contact times of interactions between individuals, is
often implemented and carried out, for instance, by partial
closure of schools and offices, and restriction of travel
\cite{Eas-10}.

In this paper, we focus on social distancing interventions to develop
a mitigation strategy that can help in reducing the number of infected
people. One way for studying these kinds of strategies is to examine
the heterogeneity in the contact times between individuals. Most
research that studies the SIR model assumes that the infection
probability is unique, which means that all individuals interact with
their neighbors in the same way. This has been disproven by several
experiments on real social networks
\cite{Catt-10,Kars-11,Steh-11}. For example, ``face-to-face''
experiments \cite{Catt-10,Steh-11} have shown that, in some cases, the
time duration of the contacts between individuals follow a power-law
distribution. This heterogeneity is called ``disorder'', and it can be
modeled using weighted networks, in which weights depend on the
normalized contact times $\omega$ of the interactions. Previous
research \cite{Buo-12,Val-13} obtains $\omega$ values from a theoretical
power law distribution with broadness $a$ (the larger the parameter
$a$, the shorter the contact times), mimicking the results of
``face-to-face'' experiments \cite{Catt-10,Steh-11}. Among other
outcomes, they found a delay in spreading of diseases as the
broadness $a$ increases \cite{Buo-12}, thereby permitting sanitary
authorities to implement earlier interventions \cite{Val-13}.

Unlike proposals of previous models, differing classes of human
interactions arise in real social networks. From the most prolonged
relationships (e.g., friendships, family, and coworkers), to the briefest
interactions (e.g., neighbors in transport and commerce), contacts between individuals
require a better distinction when modeling a social network. Motivated
by this real-world reality, we study the propagation of diseases among a population with
two coexisting types of interactions, which we distinguish by their
mean contact time. More precisely, interactions can be close or distant
with a larger or shorter mean contact time, respectively.
Each type of interaction has a
distribution of contact times governed by its own broadness $a$, which
defines the mean contact time. We use branching theory, supported by
extensive simulations, to explore a social distancing strategy that
consists of reducing only the mean contact time of distant
interactions. We propose this strategy because, generally, people are
less prone to trim their more intimate or necessary contacts, while
distant contacts are more easily controlled.

\section{Model and simulations}
\label{model}

\noindent
We construct complex networks of $N$ nodes as a substrate for the
propagation of a disease, by using the Molloy-Reed algorithm
\cite{Moll-95}. We build two types of networks with different degree
distributions. First, $P(k) = e^{-\langle k \rangle} \langle k \rangle
^k / k!$---an Erd\"os-R\'enyi network (ER)---in which $\valmed{k}$ is
the average number of neighbors of each node, and second, $P(k) = C
k^{-\lambda} e^{-k/k_c}$---a scale-free network (SF)---with
exponential cutoff $k_c$ and normalization constant $C$. ER networks
are homogeneous because their nodes have a number of neighbors mostly
around the mean value of the distribution, whereas SF networks are
heterogeneous and hence nodes have a greater amplitude in their
connectivities, with many nodes of low connectivity and only a few
nodes of high connectivity (hubs).

\begin{figure}
\centering
\includegraphics[width=0.5\textwidth]{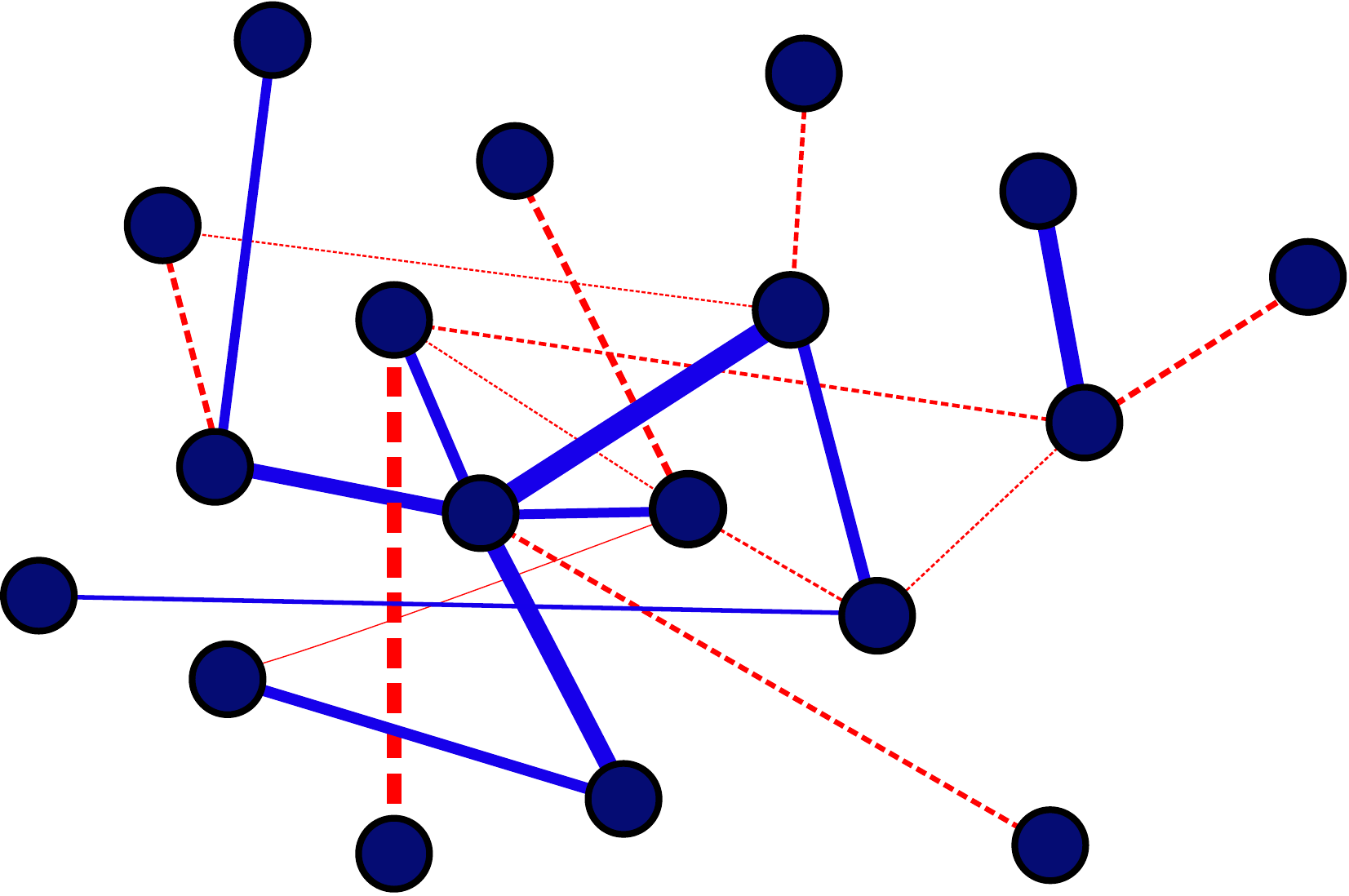}
\caption{\label{esq-modelo} Schematic of the network constructed. The
  different thicknesses of the segments represent the heterogeneity of
  the normalized contact times $\omega$ between individuals. Contact
  times belonging to the density $f_1$ (\protect\bluesolid), with $a_1
  < a_2$, are usually longer than those belonging to the complementary
  density $f_2 = 1 - f_1$ (\protect\reddash).}
\end{figure}

We use the SIR model described in Sec.~\ref{intro} to simulate the
spreading of the disease, but we assume that the infection
probability is related to the contact times between individuals, i.e.,
the more time a susceptible individual spends with an infected person,
the higher the probability they will also become infected. Thus, the
infection probability is $\beta \omega$, where a fixed $\beta$ 
represents the intrinsic virulence of the disease and $\omega$ represents the
normalized contact time between individuals. We also assume that the
contact times are heterogeneous, and hence we use a weighted network,
in which we characterize links (contacts)  by weights $\beta \omega$.
As in ``face-to-face'' experiments \cite{Catt-10}, in which contact
times follow a power law distribution, we take $\omega$ from a
theoretical distribution of contact times $P(\omega) = 1/a\omega$,
where $\omega \, \epsilon \, [e^{-a},1]$ \cite{Buo-12,Val-13}. The
parameter $a$ is called \emph{disorder intensity}, as it controls the
width of the distribution. For fixed $a$, we set $\omega = e^{-ar}$,
where $r$ is randomly selected from a uniform distribution over the
interval $[0,1]$ \cite{Brau-07}. We also separate the contacts into
two complementary parts, (i) a fraction or density $f_1$ of links with
a distribution of contact times with disorder intensity $a_1$ and (ii)
a density $f_2 = 1 - f_1$ with a distribution with disorder intensity
$a_2$, for $0 < f_1 < 1$. In Fig.~\ref{esq-modelo} we show the
density $f_1$ of interactions corresponding to the distribution with
disorder intensity $a_1$, $a_1 < a_2$ (blue continuous lines). On
average, the interactions corresponding to the density $f_1$ have
longer contact times than the ones corresponding to the density $f_2
= 1 - f_1$ (red dashed lines), indicated by the thickness of the
segments.
Our approach allows for modeling realistic populations in which different kinds
of interactions can emerge. For instance, when $a_1 < a_2$ we can
distinguish between close and distant contacts, where $a_1$ and $a_2$
are the disorder intensities of close contacts (longer average contact
times) and distant contacts (shorter average contact times),
respectively. As distant contacts are easier to control in practice,
we propose a mitigation strategy that focuses on modifying them to
reduce the scope of the disease. In Sec.~\ref{teoria} we apply
ourselves to this task.

When the propagation starts at $t=0$, all individuals are susceptible
except for one randomly-infected \emph{patient zero}. With probability
$\beta \omega$, at each time step, infected individuals propagate the
disease to their susceptible neighbors, where $\omega$ is initially
fixed and depends on the interaction between individuals. Each
infected individual recovers after a time $t_r$ since it was
infected. The spreading process ends when there are no more infected
individuals, and all are either susceptible or recovered. At this
steady state the fraction $R$ of recovered individuals for a given
value of the transmissibility $T$ indicates the extent of the disease,
since all recovered individuals were previously infected. Recall that
only realizations with $R > s_c$ are taken into account, where $s_c$
is the threshold that distinguish an epidemic ($R > s_c$) from an
outbreak ($R < s_c$).

Introducing disorder in the contact times changes the transmissibility
formula \cite{Val-13}. In our model we must account for the densities
$f_1$ or $1 - f_1$ of links that have weights $\omega$ corresponding
to the distribution with a disorder intensity of $a_1$ or $a_2$,
respectively. Then the transmissibility $T_{a_1 \, a_2}$ for a given
virulence $\beta$ and recovery time $t_r$ is 
\begin{equation}\label{e.T}
  T = T_{a_1 \, a_2} = f_1 T_{a_1} + (1 - f_1) T_{a_2},
\end{equation}
where
\begin{equation}\label{e.Ti}
  T_{a_i} = \sum_{t=1}^{t_r} \frac{(1 - \beta e^{-a_i})^t - (1 -
    \beta)^t}{a_i t}
\end{equation}
is the transmissibility of a disease in a network with a unique
distribution of contact times ($f_1 = 0$ or $f_1 = 1$), with disorder
intensity $a_i$, $i=1,2$ \cite{Val-13}. Note that the transmissibility
$T_{a_1 \, a_2}$ is a decreasing function of the intensities $a_1$ and
$a_2$, because for higher values of $a_1$ or $a_2$ the range of values
for $\omega$ allowed in each distribution of disorder expands, and
shorter contact times become more probable. Thus, the disease is less
likely to propagate. On the other hand, in the limit $a_1 \rightarrow
0$ and $a_2 \rightarrow 0$ there is no disorder, and we recover the
original (homogeneous) SIR model as $T_{a_1 \, a_2} \rightarrow T = 1 -
(1 - \beta)^{t_r}$.

\begin{figure}
\centering
\includegraphics[width=0.5\textwidth]{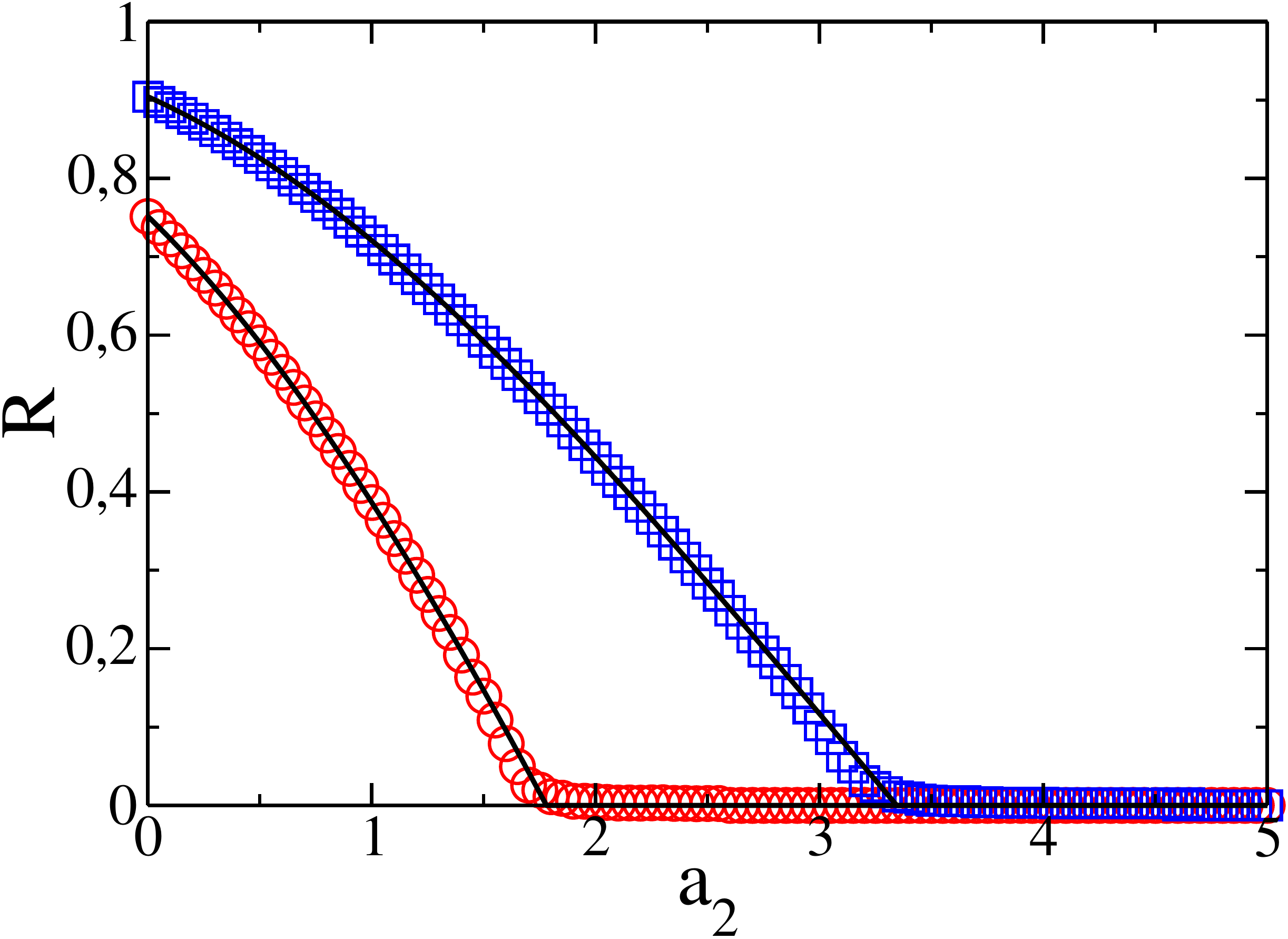}
\caption{\label{ER-sim} Fraction $R$ of recovered individuals at the
  stationary state, as a function of the disorder intensity
  $a_2$, for $\beta = 0.5$ (\protect\redcircle) and $\beta = 0.7$
  (\protect\bluesquare). Note that for each value of $\beta$ there is
  a critical value $a_{2c}$, such that the system is in a
  non-epidemic phase for $a_2 > a_{2c}$. The results of the
  simulations correspond to an ER network with $k_{\rm min} = 0$,
  $k_{\rm max} = 40$ and $\langle k \rangle = 4$, where $f_1 = 0.2$
  and $a_1 = 1$. The network size is $N = 10^5$ and $10^5$
  realizations of the simulation are performed, with $s_c = 200$. The
  black curves (\protect\blacksolid) correspond to the theoretical
  results.} 
\end{figure}

When carrying out the simulations we select, for the non-disorder
case, an infection probability $\beta$ from the epidemic phase, i.e.,
$\beta > \beta_c$ or $T > T_c$. Then, we determine whether there are
any pair of disorder intensities $(a_1,a_2)$ for which there is no
epidemic. In Fig.~\ref{ER-sim} we show the fraction $R$ of recovered
individuals as a function of the disorder $a_2$, for an ER network
with $\langle k \rangle = 4$ and different values of $\beta$, where we
fix $t_r = 1$, $f_1 = 0.2$ and $a_1 = 1$.
We can see that for both values of $\beta$, there is a critical value
$a_{2c}$ above which the system is in a non-epidemic phase. Note that
even though we have chosen a value of $\beta$ for the epidemic regime
without disorder in the network, the increasing of disorder intensity
$a_2$ reduces the spreading of the disease in the population and we
obtain a non-epidemic regime. This means that epidemics could be
reduced in size and even avoided if average contact times are
controled. In Sec.~\ref{teoria} we describe how to obtain the critical
value for the disorder intensity $a_2$ and the conditions for its
existence.

\section{Theory}
\label{teoria}

\noindent
Using the branching process formalism
\cite{New-01,New-02,New-03,Brau-07,Brau-17} we define the generating
function of the distribution $P(k)$, $G_0(x) = \sum_k P(k) x^k$, and
the generating function of the excess degree distribution $G_1(x) =
\sum_k [k P(k) / \langle k \rangle] x^{k-1}$. In Fig.~\ref{ER-sim} we
show the theoretical results for the fraction $R$ of recovered
individuals (black curves), obtained by solving the link percolation
equations $f_{\infty} = 1 - G_1(1 - p f_{\infty})$ and $P_{\infty}(p)
= 1 - G_0(1 - p f_{\infty})$, where $f_\infty$ is the probability that a
branch of links expands infinitely, $P_{\infty}$ is the fraction of
nodes in the GC, and $p$ is the fraction of links occupied on the
network. As we stated before, the SIR model can be mapped into link
percolation \cite{New-01,New-02,New-03,Brau-07,Brau-17}, thus, $R$ and
$P_{\infty}$ are equivalent. In Fig.~\ref{ER-sim} we see that the
simulation results from the SIR model with disorder present an
excellent agreement with the percolation theory. The previous
equations and the mapping between $R$ and $P_{\infty}$ apply in the
thermodynamic limit $N \to \infty$, and for locally tree-like
networks.

\begin{figure}
  \centering
  \begin{subfigure}{0.49\textwidth}
    \begin{overpic}[scale=.33]{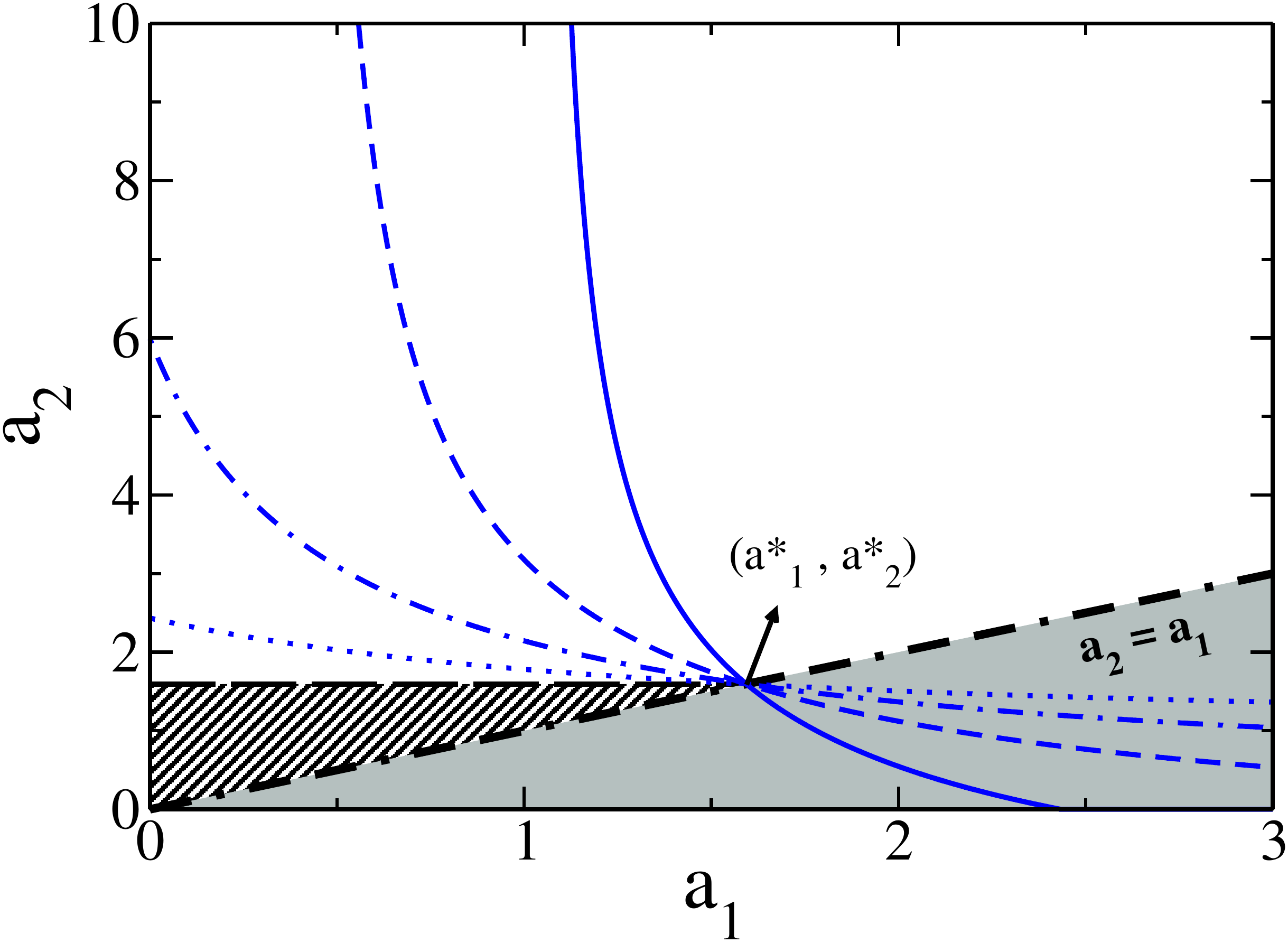}
      \put(90,65){\footnotesize\bf{(a)}}
    \end{overpic}
    \caption*{}
  \end{subfigure}
  \begin{subfigure}{0.49\textwidth}
    \begin{overpic}[scale=.33]{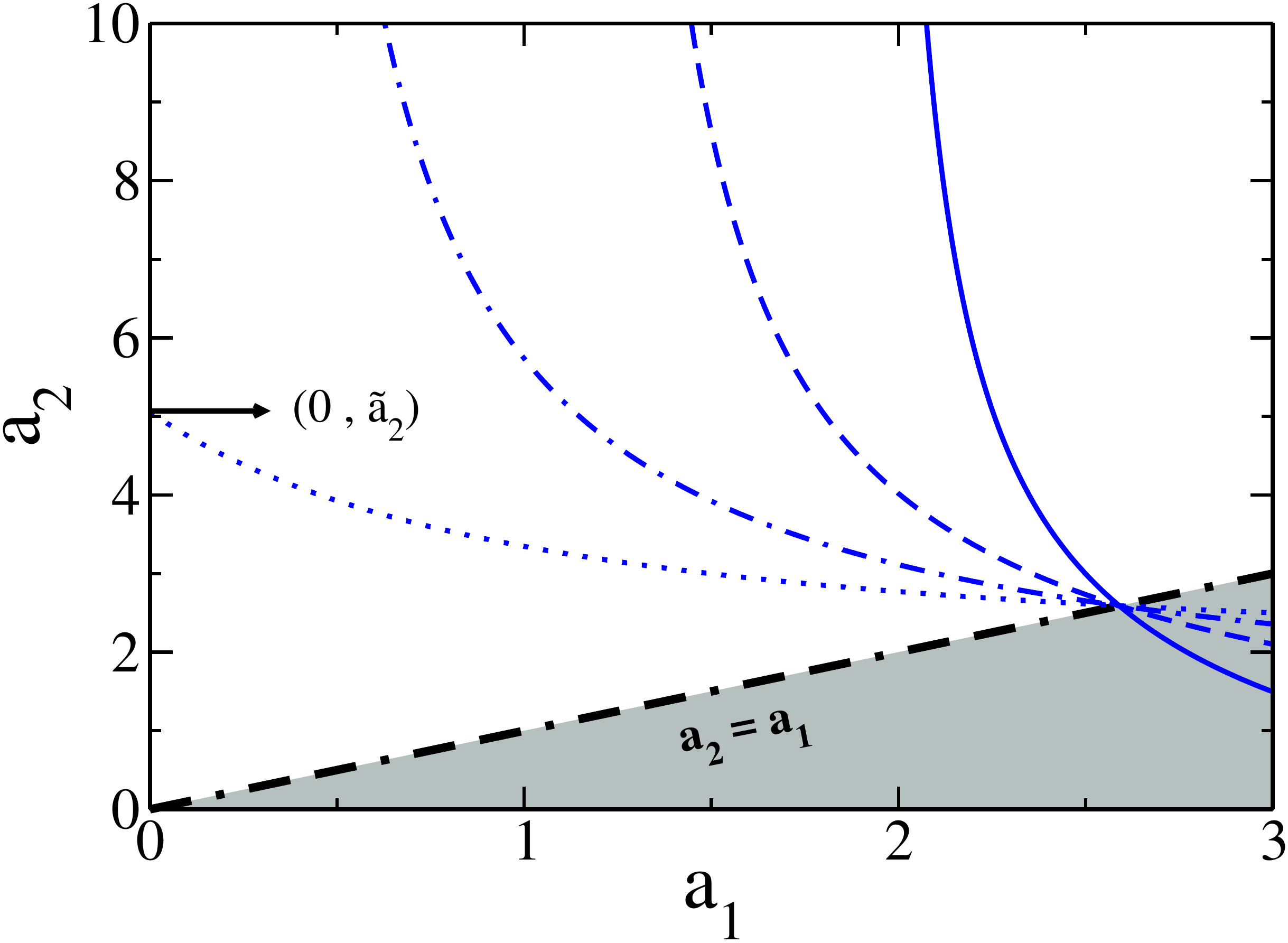}
      \put(90,65){\footnotesize\bf{(b)}}
    \end{overpic}
    \caption*{}		
  \end{subfigure}
  
  \begin{subfigure}{0.49\textwidth}
    \begin{overpic}[scale=.33]{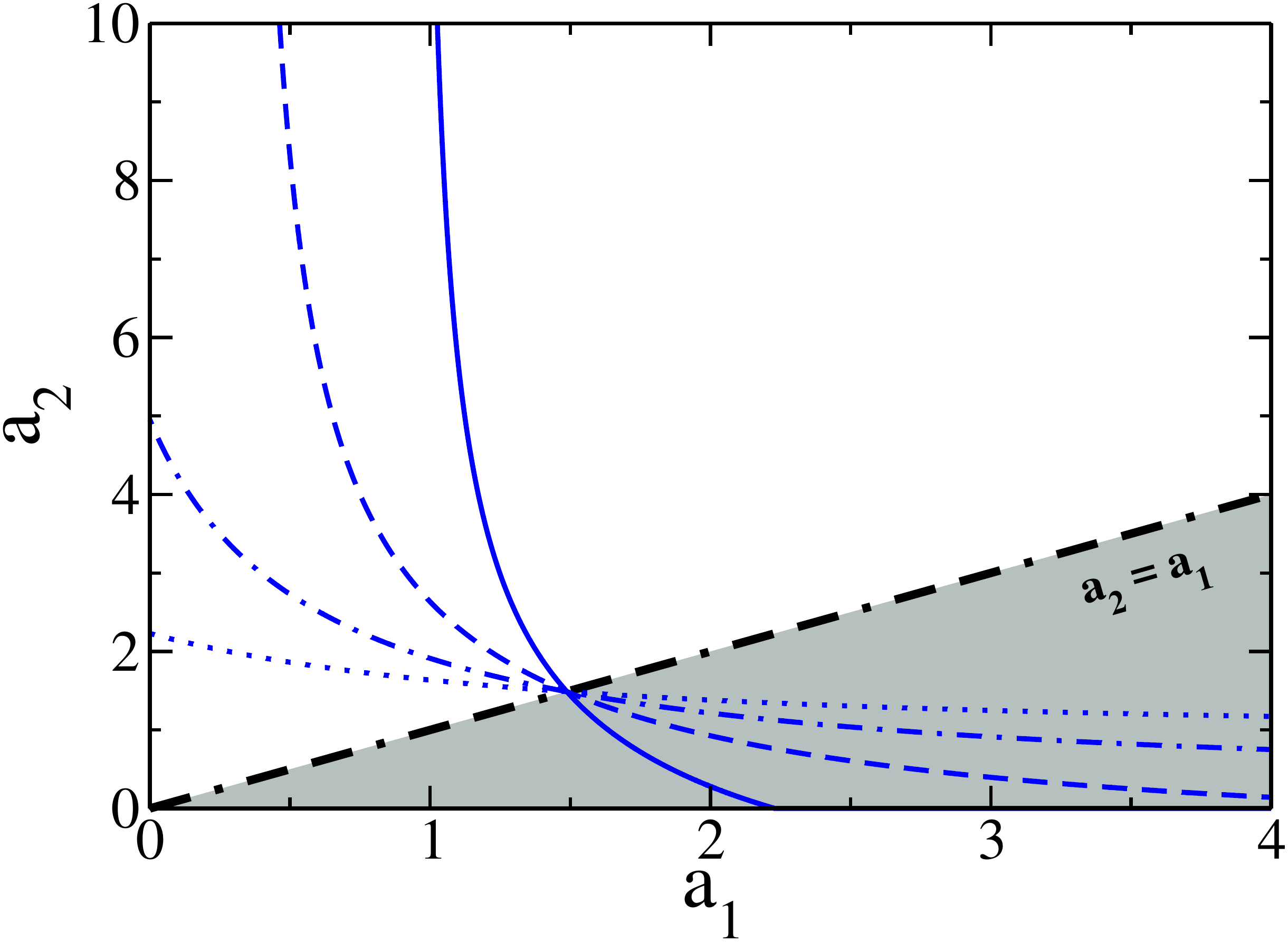}
      \put(90,65){\footnotesize\bf{(c)}}
    \end{overpic}
    \caption*{}
  \end{subfigure}
  \begin{subfigure}{0.49\textwidth}
    \begin{overpic}[scale=.33]{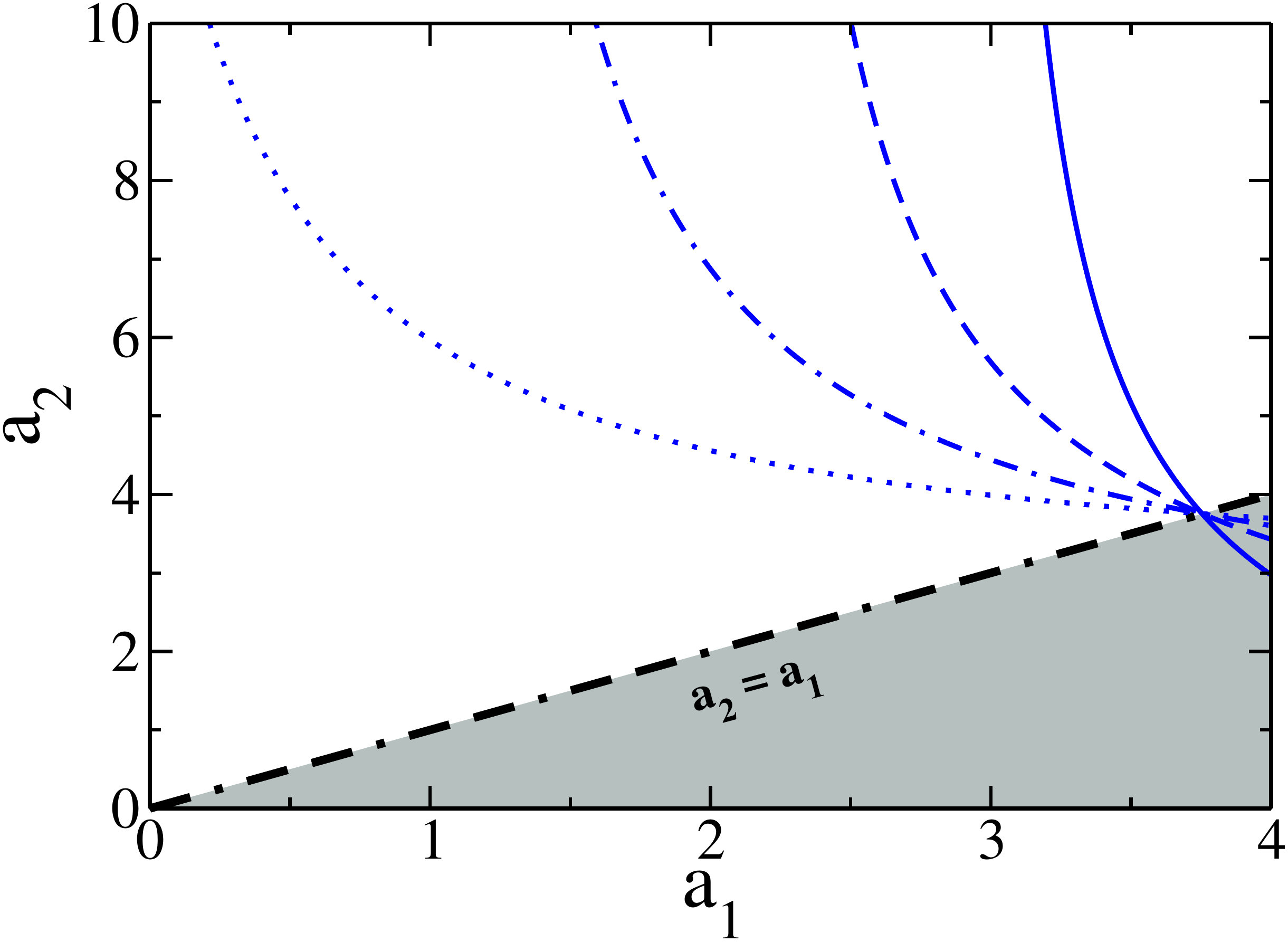}
      \put(90,65){\footnotesize\bf{(d)}}
    \end{overpic}
    \caption*{}
  \end{subfigure}
\caption{\label{esp-fases-dd-dosbetas} Phase space of the system
  projected on the $(a_1,a_2)$ plane for $t_r=1$. Each curve
  represents the critical intensity $a_{2c}$ as a function of $a_1$,
  for different densities of close contacts: $f_1 = 0.2$
  (\protect\bluedotted), $f_1 = 0.4$ (\protect\bluedashdotted), $f_1 =
  0.6$ (\protect\bluedash) and $f_1 = 0.8$ (\protect\bluesolid). Below
  each curve, the system is in an epidemic phase, while on and above
  is in an epidemic-free phase. Grey regions represents networks where
  $a_1 > a_2$, which we are not interested in. The upper figures
  correspond to an ER network with $\langle k \rangle = 4$, where (a)
  $\beta = 0.5 $ and (b) $\beta = 0.7$. The lower figures represent a
  SF network with $\lambda = 2.5$, exponential cutoff $k_c = 50$,
  where (c) $\beta = 0.25$ and (d) $\beta = 0.5$. The critical values
  of $\beta$ for a non-disordered network are $\beta_c = 0.25$ and
  $\beta_c \approx 0.13$, for degree distributions ER and SF with
  exponential cutoff respectively.}
\end{figure}

As stated in Sec.~\ref{model}, our goal is to study a mitigation
strategy for a population with both close and distant interactions, in
which we curtail the spreading of diseases by controlling the distant
contacts. If $a_1$ is the disorder intensity corresponding to
the distribution of close contacts and $a_2$ corresponds to distant
contacts, then $a_1 < a_2$. Next, we use the theoretical result from
the mapping that sets an equivalence between $T_{a_1 \, a_2}$ and $p$ to
analyze the phase space of the system, which allows us to examine our
proposed mitigation strategy. In Eq. (\ref{e.T}) we can use the
critical transmissibility $T_{a_1 \, a_{2c}} \equiv p_c = 1 / (\kappa -
1)$ to find, for $t_r = 1$,
\begin{equation}
  \label{T-a1a2-c}
  \frac{1}{\kappa -1} = f_1 \beta \frac{1 - e^{-{a_1}}}{a_1} +
  (1 - f_1) \beta \frac{1 - e^{-a_{2c}}}{a_{2c}},
\end{equation}
from which we can compute the critical intensity $a_{2c}$ for
different values of $a_1$. In Fig.~\ref{esp-fases-dd-dosbetas} we show
the phase diagram on the plane $(a_1,a_2)$ for different values of
$f_1$ and $\beta$. Because we study close and distant contacts, our
interest is focused in the region of the phase space above the
dashed-dotted line, which corresponds to networks such that $a_1 <
a_2$.
Each curve in Fig.~\ref{esp-fases-dd-dosbetas} indicates the critical
value $a_{2c}$ as a function of $a_1$ for a given density $f_1$. The
curves separate the epidemic phase (below) from the epidemic-free
phase (on and above). We also can see that in (a) there is a point
$a^*_2 = a^*_1 = a_c$ at which all the curves cross each other for
different $f_1$ values, where $a_c$ is the critical intensity for a
network with a unique disorder distribution. Starting from the $a^*_2
= a^*_1 = a_c$ point and moving away, the critical intensity $a_{2c}$
increases as $a_1$ decreases. This indicates that the longer the close
contact times, the shorter the distant contact times needed to avoid
the epidemic phase. In Fig.~\ref{esp-fases-dd-dosbetas}(b) we show
that $a_1$ can even go to zero, which means that the close contact
times can be as long as possible. In this limit we see that $a_{2c}$
converges to a finite value $\tilde{a}_2$. Using Eq.~(\ref{T-a1a2-c})
we obtain an expression for $\tilde{a}_2$,
\begin{equation}
  T_c = f_1 \beta + (1 - f_1) \beta
  \frac{1 - e^{-\tilde{a}_{2}}}{\tilde{a}_{2}}.
  \label{free-epidemic}
\end{equation}
Using Eq.~(\ref{free-epidemic}) we find that $\tilde{a}_2$ exists if
$f_1 < T_c / \beta \equiv \tilde{f}_1$, otherwise the close contacts
cause the system to be in an epidemic phase for any value of $a_2$,
which means that $\tilde{a}_2$ does not exist. In this case, when $f_1
> \tilde{f}_1$, $a_{2c} \rightarrow \infty$ as $a_1 \rightarrow
a_{1m}$ [see Eq.~(\ref{T-a1a2-c})]. Thus, distant contact times are
equal to zero, and because the disease cannot pass through these
contacts its corresponding transmissibility is also zero. The
resulting expression for $a_1 = a_{1m}$ is then
\begin{equation}
  T_c = f_1 \beta \frac{1 - e^{-a_{1m}}}{a_{1m}}.
  \label{a1c-min}
\end{equation}
Since there is no critical value $a_{2c}$ for $a_1 < a_{1m}$, the
disease is always in an epidemic phase.

Note that there is a region of the phase space (striped region) in 
which the disease is in an epidemic phase for all $f_1$ values [see
Fig.~\ref{esp-fases-dd-dosbetas}(a)]. This region corresponds to the
epidemic phase for $f_1 = 0$, i.e., when there is only one type of
contacts in the network. Then, it is characterized by $T_{a2} > T_c =
1 / (\kappa - 1)$.

\begin{figure}
  \centering
  \begin{subfigure}{0.49\textwidth}
    \begin{overpic}[scale=.33]{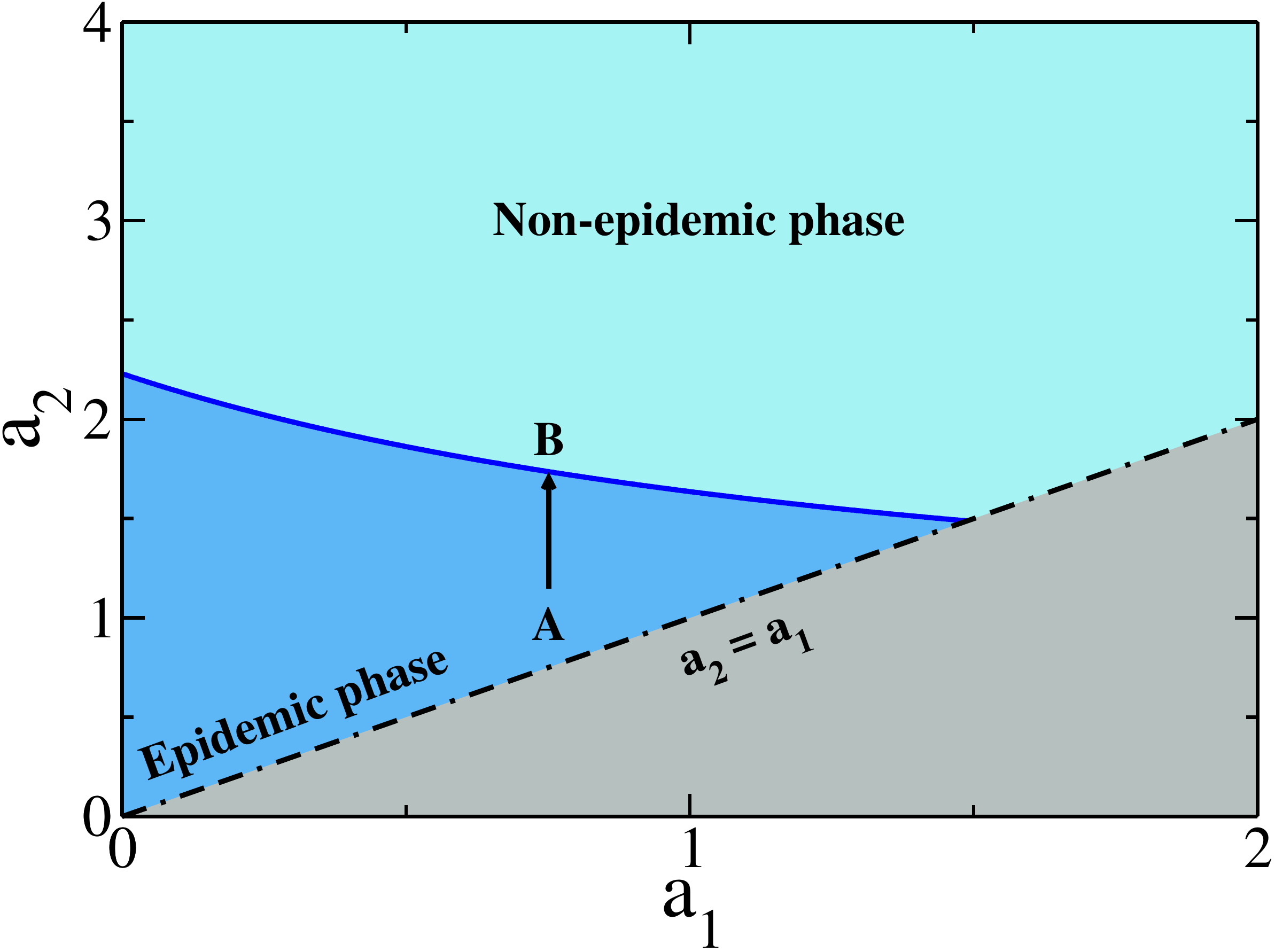}
      \put(90,67){\footnotesize\bf{(a)}}
    \end{overpic}
    \caption*{}
  \end{subfigure}
  \begin{subfigure}{0.49\textwidth}
    \begin{overpic}[scale=.33]{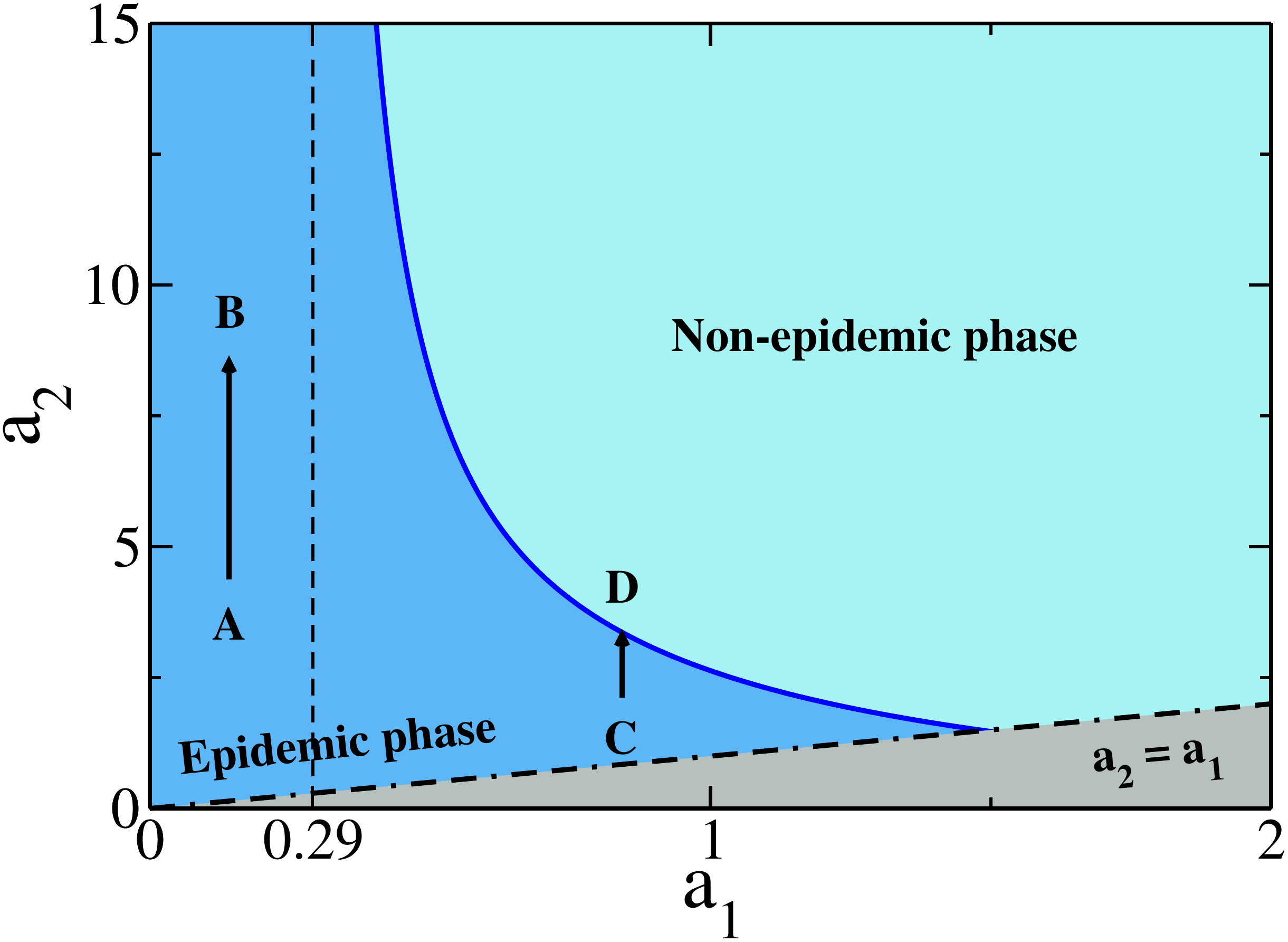}
      \put(90,65){\footnotesize\bf{(b)}}
    \end{overpic}
    \caption*{}
  \end{subfigure}
\caption{\label{estrategia} Schematic of the proposed strategy to halt
  the spreading of a disease with virulence $\beta = 0.25$. In (a) we
  show the case $f_1 < \tilde{f}_1 = T_c / \beta$ ($f_1 = 0.2$), where
  $\tilde{a}_2$ exists for $a_1 = 0$. Then, starting from any point
  $A$ in the epidemic phase, by increasing $a_2$ we can reach the
  critical point in $B$. The opposite case ($f_1 > \tilde{f}_1$) is
  represented in (b) ($f_1 = 0.6$), which shows the same behavior than
  in (a) from point $C$ to the critical point in $D$, for the case in
  which $a_1$ is originally greater than or equal to the minimum value
  $a_{1m}$, corresponding to $a_{2c} \rightarrow \infty$. Here $a_{1m}
  = 0.29$. The results correspond to a SF network with $\lambda = 2.5$
  and exponential cutoff $k_c = 50$.}
\end{figure}

We use these results to construct a distancing strategy for reducing
the impact of a disease in a population with close and distant
contacts, by controlling the duration of distant contact
times. Suppose that the distribution of contact times has original
disorder intensities $a_1$ and $a_2$ such that the system is in an
epidemic phase. Then, if we assume that close contacts are a minor
portion of the total ($f_1 < \tilde{f}_1 = T_c / \beta$), we can
increase the intensity $a_2$ to a critical point, hence reaching a
non-epidemic phase independent of the original intensities [see
  Fig.~\ref{estrategia}(a)]. When $f_1 > \tilde{f}_1$, the original
value of the disorder intensity of close contacts determines whether
we can reach the non-epidemic phase [see Fig.~\ref{estrategia}(b)]. In
this case, when $a_1 < a_{1m}$ the non-epidemic phase cannot be
reached by simply increasing $a_2$.

We also observe that, with fixed $\beta$, the critical values
obtained for ER networks are lower than the ones obtained for SF
networks with an exponential cutoff. We can see this result by comparing
Figs.~\ref{esp-fases-dd-dosbetas} (a) and (d). In homogeneous (ER)
networks individuals have, on average, the same number of
neighbors. Thus, there is a limit on the speed at which the disease
can propagate. In contrast, the presence of hubs in heterogeneous (SF)
networks causes a rapid propagation of the disease
once they become infected. Therefore, these networks require higher
disorder intensities (or shorter contact times) to reach a
non-epidemic phase than those required in ER networks. Note also that
the intrinsic virulence of the disease $\beta$ modifies these critical
values.

In Figs.~\ref{esp-fases-dd-dosbetas}(a) and
\ref{esp-fases-dd-dosbetas}(b), and in
Figs.~\ref{esp-fases-dd-dosbetas}(c) and
\ref{esp-fases-dd-dosbetas}(d), we show that when $\beta$ increases
the disease becomes more aggressive, spreads more rapidly, critical
intensities increase, and the epidemic phase of the disease widens.

\begin{figure}
  \centering
  \begin{subfigure}{0.49\textwidth}
    \begin{overpic}[scale=.33]{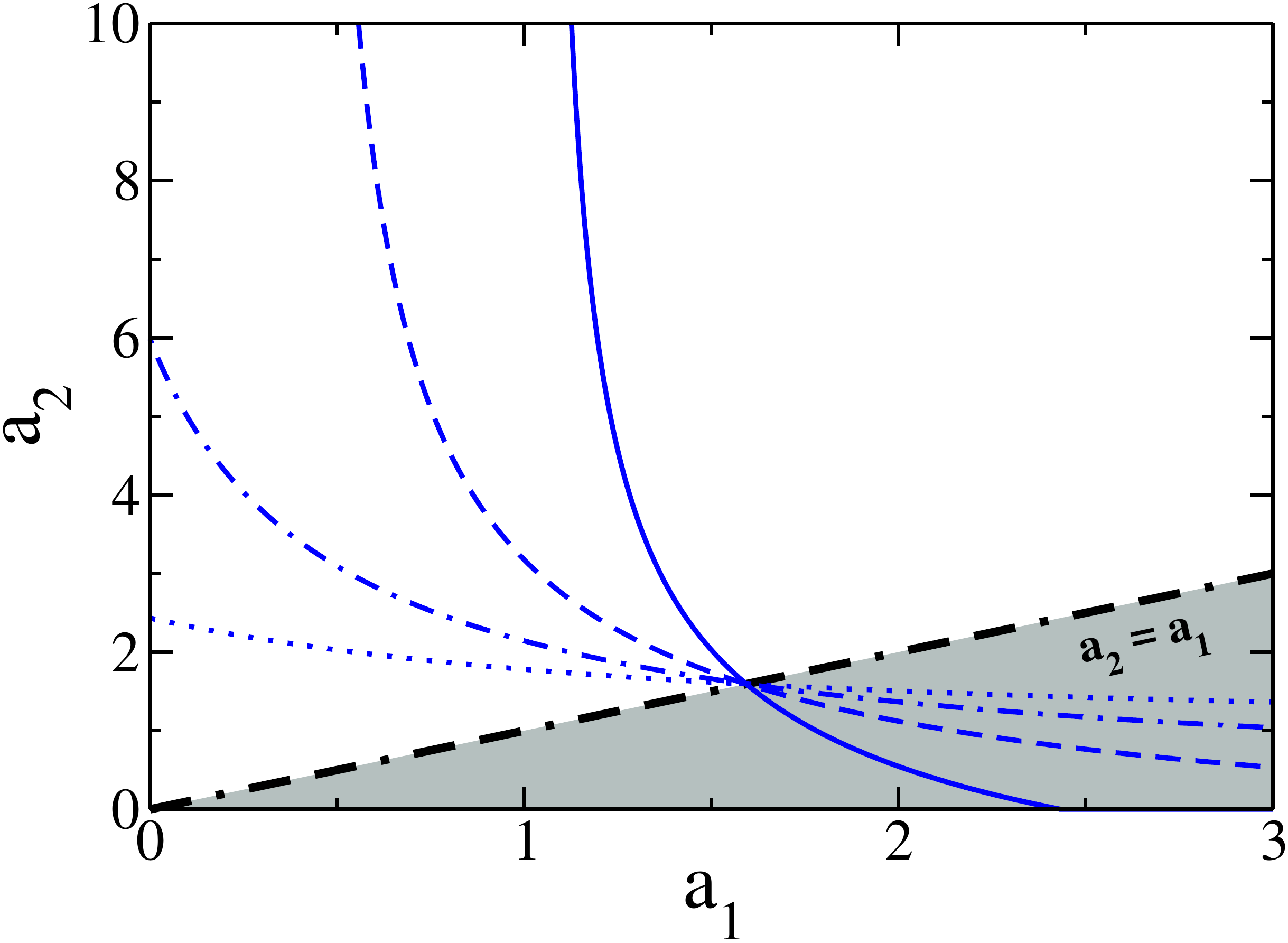}
      \put(90,65){\footnotesize\bf{(a)}}
    \end{overpic}
    \caption*{}
  \end{subfigure}
  \begin{subfigure}{0.49\textwidth}
    \begin{overpic}[scale=.33]{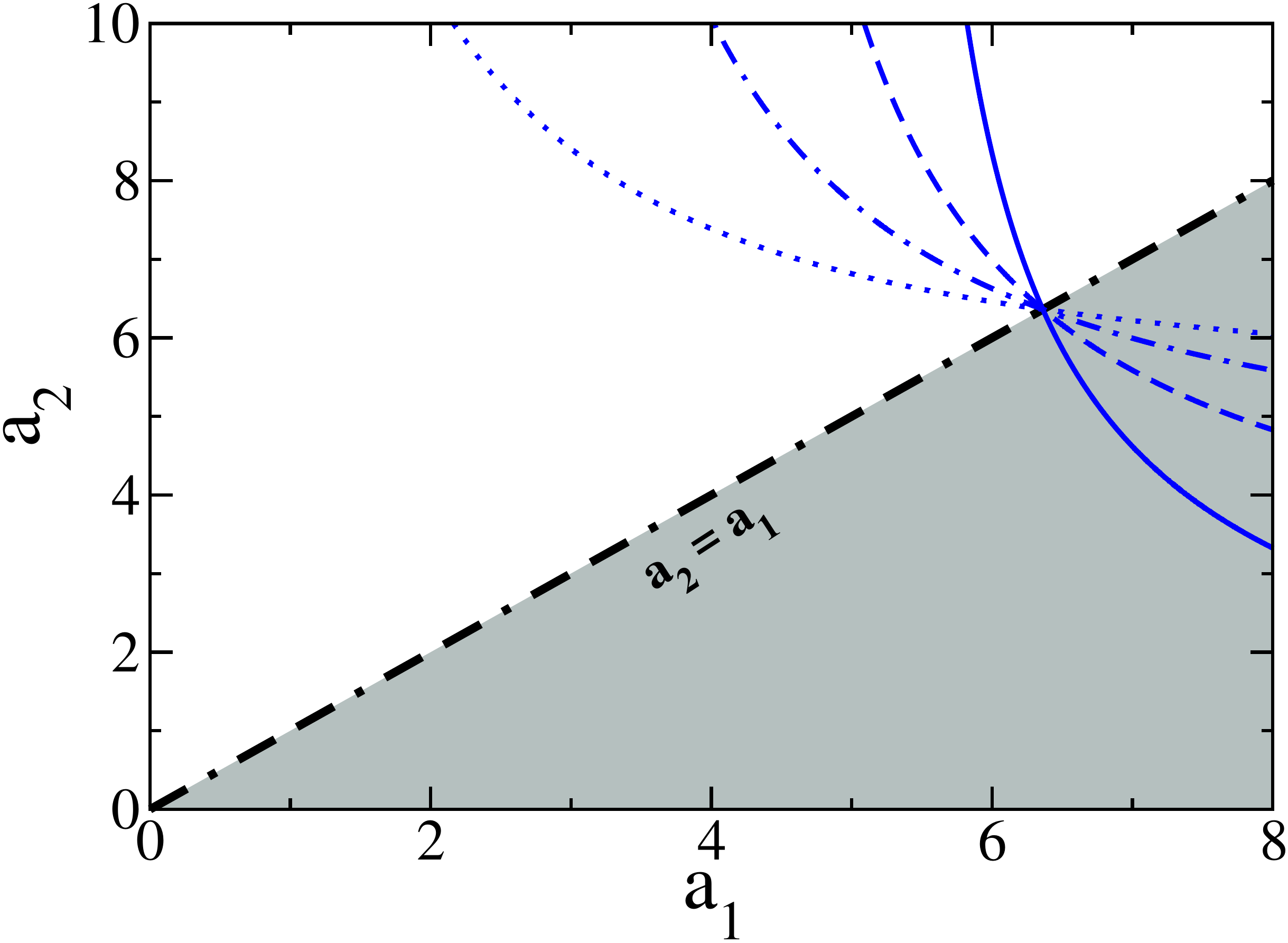}
      \put(14,65){\footnotesize\bf{(b)}}
    \end{overpic}
    \caption*{}		
  \end{subfigure}
  
  \begin{subfigure}{0.49\textwidth}
    \begin{overpic}[scale=.33]{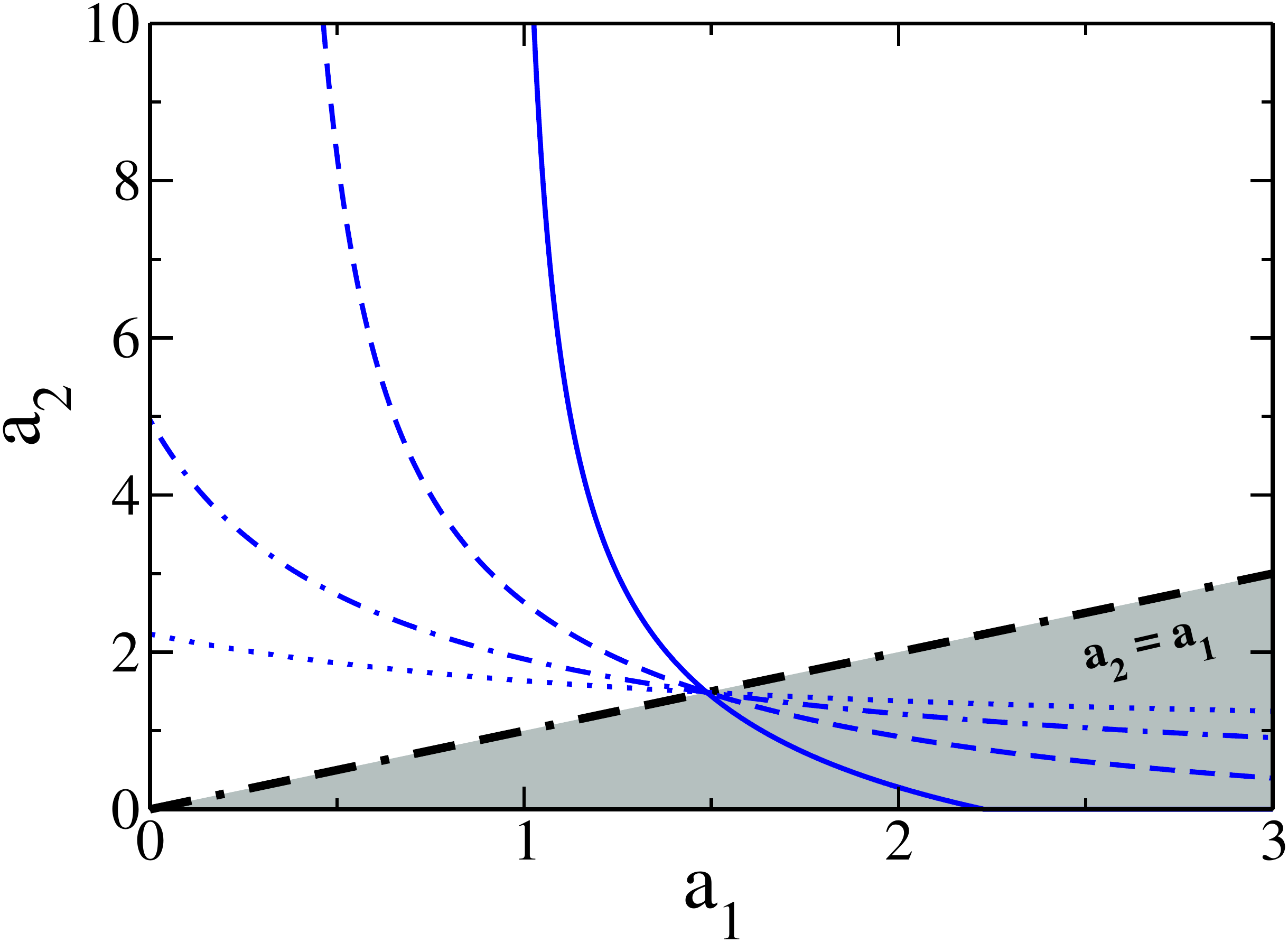}
      \put(90,65){\footnotesize\bf{(c)}}
    \end{overpic}
    \caption*{}
  \end{subfigure}
  \begin{subfigure}{0.49\textwidth}
    \begin{overpic}[scale=.33]{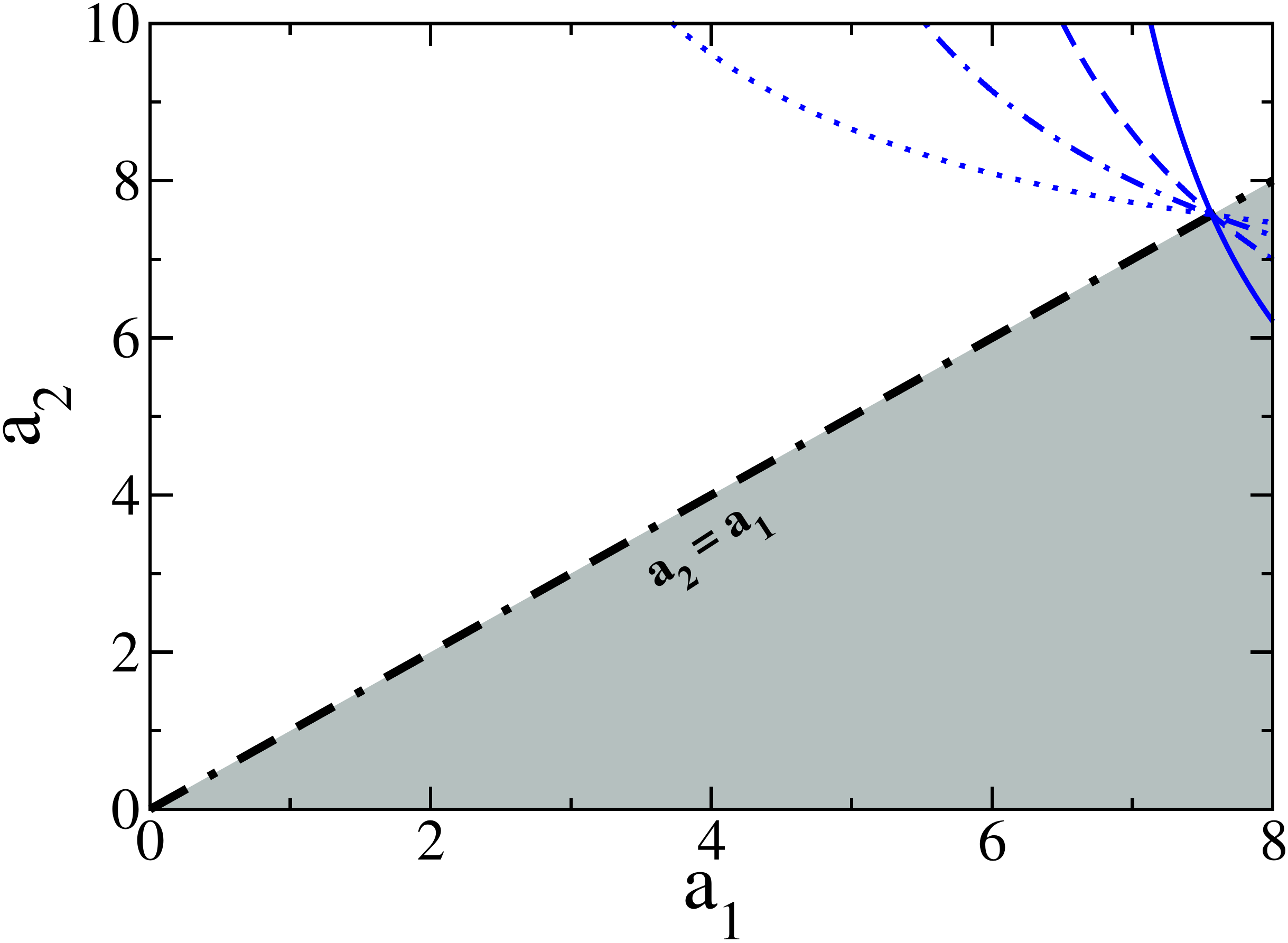}
      \put(14,65){\footnotesize\bf{(d)}}
    \end{overpic}
    \caption*{}
  \end{subfigure}
\caption{\label{esp-fases-dd-dostr} Phase space of the system
  projected on the $(a_1,a_2)$ plane, for different densities of close
  contacts: $f_1 = 0.2$ (\protect\bluedotted), $f_1 = 0.4$
  (\protect\bluedashdotted), $f_1 = 0.6$ (\protect\bluedash) and $f_1
  = 0.8$ (\protect\bluesolid). The upper figures correspond to an ER
  network with $\langle k \rangle = 4$ and $\beta = 0.5$, for (a) $t_r
  = 1$ and (b) $t_r = 5$. The lower figures correspond to a SF network
  with $\lambda = 2.5$, exponential cutoff $k_c = 50$, and $\beta =
  0.25$, for (c) $t_r = 1$ and (d) $t_r = 5$. Note that the critical
  intensities take greater values to counter the increase of the
  recovery times.}
\end{figure}

Finally, we generalize the analysis for larger recovery times ($t_r >
1$). In Fig.~\ref{esp-fases-dd-dostr} we show the phase space obtained
from Eqs.~(\ref{e.T}) and (\ref{e.Ti}) for $t_r = 5$. This could
represent the situation of a disease such as the flu, which has a mean
recovery time of five days. Also, in Fig.~\ref{esp-fases-dd-dostr} we
compare these results with the $t_r = 1$ case.
Note that results for different recovery times $t_r$ do not
qualitatively differ. However, for fixed $f_1$, the epidemic phase
becomes wider when $t_r$ increases. This is because the infected
individuals have more time to propagate the disease, and thus the
contact times must be shorter (or have larger disorder intensities) to
move the disease to a non-epidemic phase. The recovery time is an
important factor that needs to be accounted for in the implementation
of our epidemic-avoiding strategy, and it varies depending on the
type of disease.

\section{Analysis for the distribution $P'(\omega) = 1/(a'_1\omega^{1.6})$}

\noindent
Some ``face-to-face'' experiments have studied the contact behavior of
individuals at conference-like reunions. The duration of these
interactions is accurately reflected by a distribution $P'(\omega)
\propto \omega^{-1.6}$ \cite{Catt-10,Steh-11}. For a more realistic
analysis, we include this distribution in our model with a density
$f_1$ of close contacts. We make this selection because individuals at
conferences usually spend most of their time with the same group
of people, a contact pattern that we define to be close. We compare
our previous results with those produced by this new distribution,
strictly defined by $P'(\omega) = 1/(a'_1\omega^{1.6})$, where
$\omega \, \epsilon \,[(1 +   0.6a'_1)^{-5/3},1]$ and $a'_1$ is the
disorder intensity.

\begin{figure}
  \centering
  \begin{subfigure}{0.49\textwidth}
    \begin{overpic}[scale=.33]{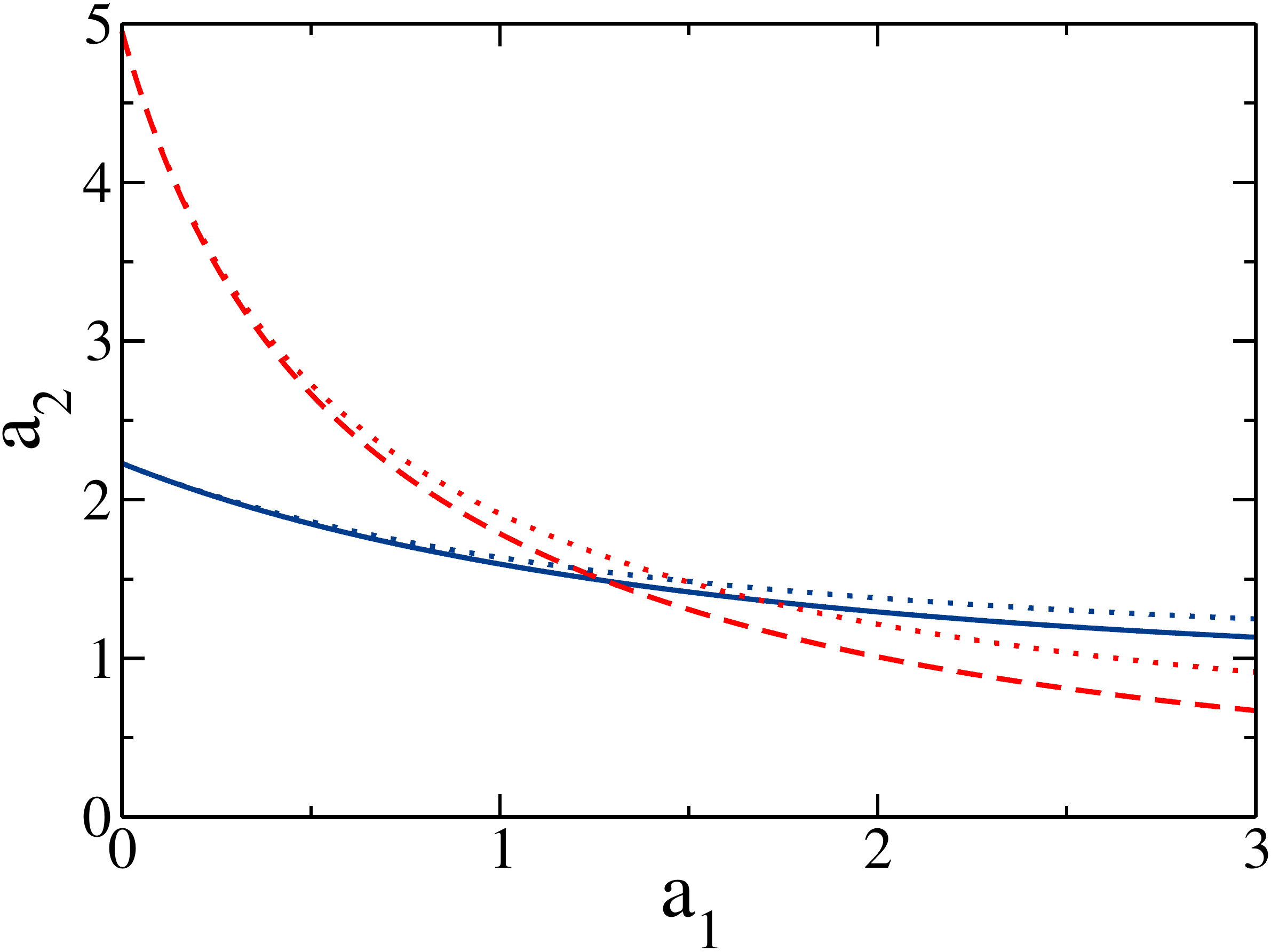}
      \put(90,67){\footnotesize\bf{(a)}}
    \end{overpic}
    \caption*{}
  \end{subfigure}
  \begin{subfigure}{0.49\textwidth}
    \begin{overpic}[scale=.33]{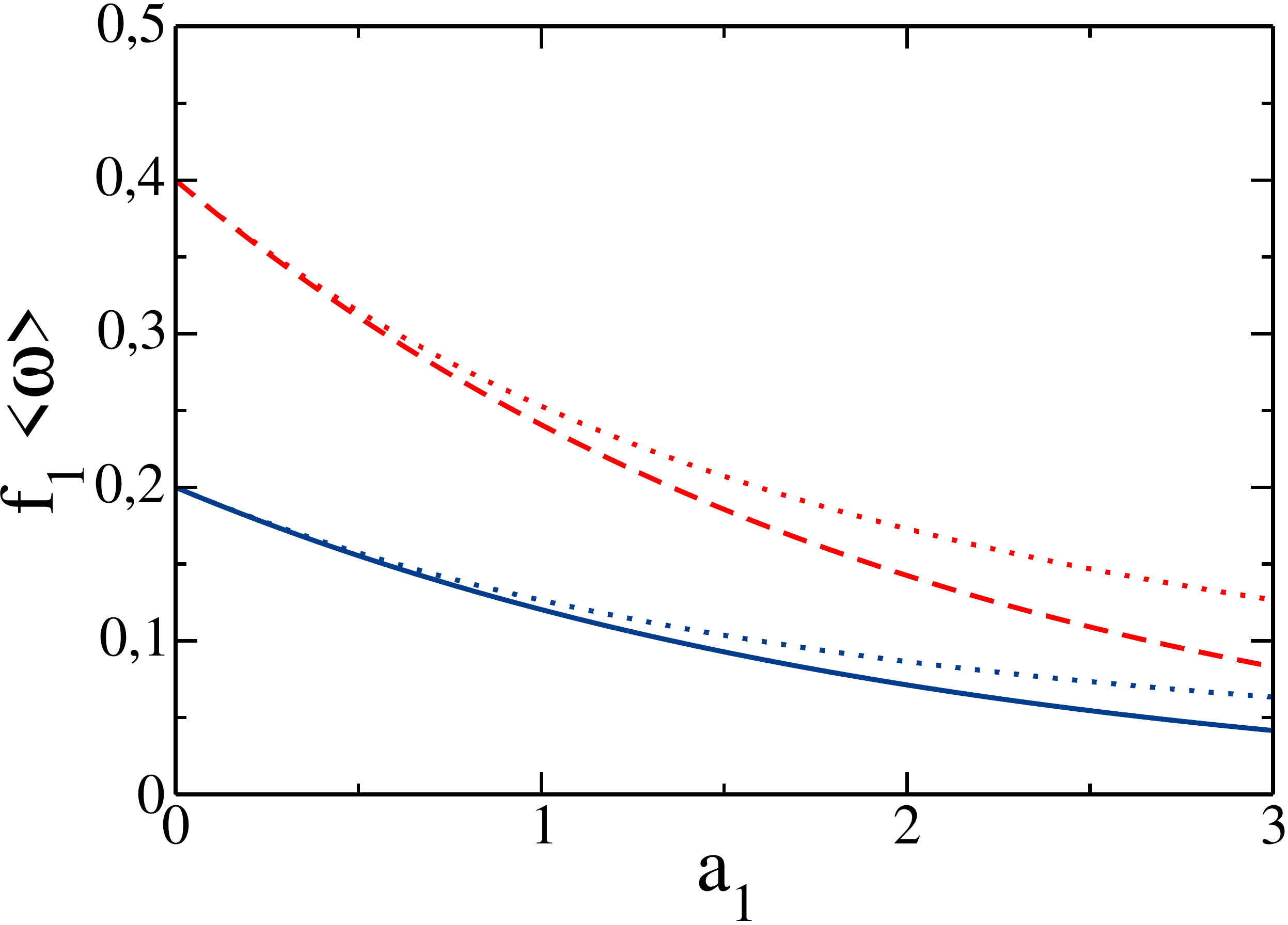}
      \put(90,64){\footnotesize\bf{(b)}}
    \end{overpic}
    \caption*{}		
  \end{subfigure}
\caption{\label{comparacion-dists} (a) Critical intensities $a_{2c}$
  and (b) average normalized contact times $f_1 \valmed{\omega}$, as
  functions of $a_1$, for different densities of close contacts
  distributed according to $P'(\omega) =  1/(a'_1 \omega^{1.6})$:
  $f_1 = 0.2$ (\protect\bluesolid) and $f_1 = 0.4$
  (\protect\reddash). The dotted lines are the corresponding results
  previously obtained for the distribution of close contacts
  $P(\omega) = 1 / (a_1 \omega)$. Disorder intensity $a_1$ is such
  that the minimum values of $\omega$ coincide for both
  distributions. Results shown in (a) correspond to a SF network with
  $\lambda = 2.5$, exponential cutoff $k_c = 50$, and for $\beta =
  0.25$ and $t_r = 1$. Note that in (b), for a fixed value of $f_1$,
  the difference between average contact times increases with $a_1$,
  i.e., when the range of allowed contact times becomes wider.}
\end{figure}

As we stated before, now we work with a population in which a density
$f_1$ of the interactions has a contact time distribution $P'(\omega)
= 1/(a'_1 \omega^{1.6})$ and the density $f_2 = 1 - f_1$ is distributed
according to $P(\omega) = 1/(a_2 \omega)$. We want to compare this
scenario with the previously studied case, which only differs in that
the the distribution of the density $f_1$ of contact times is
$P(\omega) = 1 / (a_1 \omega)$. In order to accurately compare these
distributions, the normalized contact time ranges must be the same for
both and thus, the minimum $\omega$ values must be equal. This yields
$(1 + 0.6a'_1)^{-5/3} = e^{-a_1}$ and gives a relation between the
disorder intensities $a_1$ and $a'_1$. For a fixed value of $a_1$, we
compute the corresponding value for $a'_1$ and use these two
intensities to obtain the critical values $a_{2c}$ for each
case. Then, we plot $a_{2c}$ as a function of $a_1$ for both cases
[see Fig.~\ref{comparacion-dists}(a)]. This allows a comparison
of the results when both distributions have the same range of
normalized contact times.
We can see that for the distribution $P'(\omega)$, the critical values
$a_{2c}$ are smaller than those that were previously obtained for
$P(\omega)$, which means that the disease spreads more easily under
the distribution $P(\omega)$. We can understand this if we observe
Fig.~\ref{comparacion-dists} (b), where we show a comparison between
the average normalized contact times $f_1 \valmed{\omega}'$ and $f_1
\valmed{\omega}$, corresponding to the density $f_1$ of contacts
distributed according to $P'(\omega)$ and $P(\omega)$,
respectively. For any $a_1$ value $f_1 \valmed{\omega}' < f_1
\valmed{\omega}$, which means that the disease is less likely to
propagate through interactions when the contact times are distributed
according to $P'(\omega)$, the more realistic distribution of contact
times that we defined from the experiments.

\section{Conclusions}
\label{conclusiones}

\noindent
In this paper, we study the SIR model for disease spreading over a
disordered complex network, in which two types of interactions are
defined: close and distant contacts, with larger and shorter mean
contact times, respectively. We propose a mitigation strategy
consisting in reducing the average contact time of distant
interactions (by increasing $a_2$ in the model). We find that the
strategy is more effective for smaller densities $f_1$ of close
contacts, as the disease is more likely to propagate through
them. Also, there is a threshold density $\tilde{f}_1 = T_c / \beta$
of close contacts below which the strategy can prevent the system to
enter in an epidemic phase, regardless of the average contact time of
close interactions. Using a distribution of close contact times
$P'(\omega) = 1/(a'_1\omega^{1.6})$ that adjusts better with some
experimental results, we find that the propagation decreases and it is
easier to reduce the impact of a disease than when using the
theoretical distribution $P(\omega) = 1 / (a_1 \omega)$.

The analysis carried out here can be extended to interconnected
networks, where each network represents a different environment in
which interactions take place. As differing networks can have their
own degree distribution, this could be an approach for making our close/distant
interaction model more realistic and broadly applicable. Since it is
well known that such
interconnected systems accelerate spreading processes,  it is
fundamental, in these cases, to find ways to halt or slow them down.

\acknowledgments

\noindent
We acknowledge UNMdP and CONICET (PIP 00443/2014) for financial
support. CELR and IAP acknowledges CONICET for financial support. Work
at Boston University is supported by NSF Grants PHY-1505000 and by DTRA
Grant HDTRA1-14-1-0017. We gratefully thank Mat\'ias A. Di Muro for
useful comments.


\begin{thebibliography}{38}%
\makeatletter
\providecommand \@ifxundefined [1]{%
 \@ifx{#1\undefined}
}%
\providecommand \@ifnum [1]{%
 \ifnum #1\expandafter \@firstoftwo
 \else \expandafter \@secondoftwo
 \fi
}%
\providecommand \@ifx [1]{%
 \ifx #1\expandafter \@firstoftwo
 \else \expandafter \@secondoftwo
 \fi
}%
\providecommand \natexlab [1]{#1}%
\providecommand \enquote  [1]{``#1''}%
\providecommand \bibnamefont  [1]{#1}%
\providecommand \bibfnamefont [1]{#1}%
\providecommand \citenamefont [1]{#1}%
\providecommand \href@noop [0]{\@secondoftwo}%
\providecommand \href [0]{\begingroup \@sanitize@url \@href}%
\providecommand \@href[1]{\@@startlink{#1}\@@href}%
\providecommand \@@href[1]{\endgroup#1\@@endlink}%
\providecommand \@sanitize@url [0]{\catcode `\\12\catcode `\$12\catcode
  `\&12\catcode `\#12\catcode `\^12\catcode `\_12\catcode `\%12\relax}%
\providecommand \@@startlink[1]{}%
\providecommand \@@endlink[0]{}%
\providecommand \url  [0]{\begingroup\@sanitize@url \@url }%
\providecommand \@url [1]{\endgroup\@href {#1}{\urlprefix }}%
\providecommand \urlprefix  [0]{URL }%
\providecommand \Eprint [0]{\href }%
\providecommand \doibase [0]{http://dx.doi.org/}%
\providecommand \selectlanguage [0]{\@gobble}%
\providecommand \bibinfo  [0]{\@secondoftwo}%
\providecommand \bibfield  [0]{\@secondoftwo}%
\providecommand \translation [1]{[#1]}%
\providecommand \BibitemOpen [0]{}%
\providecommand \bibitemStop [0]{}%
\providecommand \bibitemNoStop [0]{.\EOS\space}%
\providecommand \EOS [0]{\spacefactor3000\relax}%
\providecommand \BibitemShut  [1]{\csname bibitem#1\endcsname}%
\let\auto@bib@innerbib\@empty
%</preamble>
\bibitem [{\citenamefont {Anderson}\ and\ \citenamefont
  {May}(1992)}]{Ander-92}%
  \BibitemOpen
  \bibfield  {author} {\bibinfo {author} {\bibfnamefont {R.~M.}\ \bibnamefont
  {Anderson}}\ and\ \bibinfo {author} {\bibfnamefont {R.~M.}\ \bibnamefont
  {May}},\ }\href@noop {} {\emph {\bibinfo {title} {{Infectious Diseases of
  Humans: Dynamics and Control}}}}\ (\bibinfo  {publisher} {Oxford University
  Press, Oxford},\ \bibinfo {year} {1992})\BibitemShut {NoStop}%
\bibitem [{\citenamefont {Johnson}\ and\ \citenamefont
  {Mueller}(2002)}]{Joh-02}%
  \BibitemOpen
  \bibfield  {author} {\bibinfo {author} {\bibfnamefont {N.~P. A.~S.}\
  \bibnamefont {Johnson}}\ and\ \bibinfo {author} {\bibfnamefont
  {J.}~\bibnamefont {Mueller}},\ }\href {\doibase 10.1353/bhm.2002.0022}
  {\bibfield  {journal} {\bibinfo  {journal} {Bull. Hist. Med.}\ }\textbf
  {\bibinfo {volume} {76}},\ \bibinfo {pages} {105} (\bibinfo {year}
  {2002})}\BibitemShut {NoStop}%
\bibitem [{\citenamefont {Fraser}\ \emph {et~al.}(2009)\citenamefont {Fraser},
  \citenamefont {Donnelly}, \citenamefont {Cauchemez}, \citenamefont {Hanage},
  \citenamefont {Van~Kerkhove}, \citenamefont {Hollingsworth}, \citenamefont
  {Griffin}, \citenamefont {Baggaley}, \citenamefont {Jenkins}, \citenamefont
  {Lyons} \emph {et~al.}}]{Fras-09}%
  \BibitemOpen
  \bibfield  {author} {\bibinfo {author} {\bibfnamefont {C.}~\bibnamefont
  {Fraser}}, \bibinfo {author} {\bibfnamefont {C.~A.}\ \bibnamefont
  {Donnelly}}, \bibinfo {author} {\bibfnamefont {S.}~\bibnamefont {Cauchemez}},
  \bibinfo {author} {\bibfnamefont {W.~P.}\ \bibnamefont {Hanage}}, \bibinfo
  {author} {\bibfnamefont {M.~D.}\ \bibnamefont {Van~Kerkhove}}, \bibinfo
  {author} {\bibfnamefont {T.~D.}\ \bibnamefont {Hollingsworth}}, \bibinfo
  {author} {\bibfnamefont {J.}~\bibnamefont {Griffin}}, \bibinfo {author}
  {\bibfnamefont {R.~F.}\ \bibnamefont {Baggaley}}, \bibinfo {author}
  {\bibfnamefont {H.~E.}\ \bibnamefont {Jenkins}}, \bibinfo {author}
  {\bibfnamefont {E.~J.}\ \bibnamefont {Lyons}},  \emph {et~al.},\ }\href
  {\doibase 10.1126/science.1176062} {\bibfield  {journal} {\bibinfo  {journal}
  {Science}\ }\textbf {\bibinfo {volume} {324}},\ \bibinfo {pages} {1557}
  (\bibinfo {year} {2009})}\BibitemShut {NoStop}%
\bibitem [{\citenamefont {Merler}\ \emph {et~al.}(2015)\citenamefont {Merler},
  \citenamefont {Ajelli}, \citenamefont {Fumanelli}, \citenamefont {Gomes},
  \citenamefont {Pastore~y Piontti}, \citenamefont {Rossi}, \citenamefont
  {Chao}, \citenamefont {Longini}, \citenamefont {Halloran},\ and\
  \citenamefont {Vespignani}}]{Mer-15}%
  \BibitemOpen
  \bibfield  {author} {\bibinfo {author} {\bibfnamefont {S.}~\bibnamefont
  {Merler}}, \bibinfo {author} {\bibfnamefont {M.}~\bibnamefont {Ajelli}},
  \bibinfo {author} {\bibfnamefont {L.}~\bibnamefont {Fumanelli}}, \bibinfo
  {author} {\bibfnamefont {M.~F.~C.}\ \bibnamefont {Gomes}}, \bibinfo {author}
  {\bibfnamefont {A.}~\bibnamefont {Pastore~y Piontti}}, \bibinfo {author}
  {\bibfnamefont {L.}~\bibnamefont {Rossi}}, \bibinfo {author} {\bibfnamefont
  {D.~L.}\ \bibnamefont {Chao}}, \bibinfo {author} {\bibfnamefont {I.~M.}\
  \bibnamefont {Longini}}, \bibinfo {author} {\bibfnamefont {M.~E.}\
  \bibnamefont {Halloran}}, \ and\ \bibinfo {author} {\bibfnamefont
  {A.}~\bibnamefont {Vespignani}},\ }\href {\doibase
  10.1016/S1473-3099(14)71074-6} {\bibfield  {journal} {\bibinfo  {journal}
  {Lancet Infect. Dis.}\ }\textbf {\bibinfo {volume} {15}},\ \bibinfo {pages}
  {204} (\bibinfo {year} {2015})}\BibitemShut {NoStop}%
\bibitem [{\citenamefont {Fox}(2019)}]{NBC-19}%
  \BibitemOpen
  \bibfield  {author} {\bibinfo {author} {\bibfnamefont {M.}~\bibnamefont
  {Fox}},\ }\href
  {https://www.nbcnews.com/health/health-news/new-york-fighting-its-worst-outbreak-measles-decades-n955891}
  {\enquote {\bibinfo {title} {New york is fighting its worst outbreak of
  measles in decades},}\ } (\bibinfo {year} {2019})\BibitemShut {NoStop}%
\bibitem [{\citenamefont {{A. E. van den Bogaard}}\ and\ \citenamefont
  {Stobberingh}(2000)}]{Vand-00}%
  \BibitemOpen
  \bibfield  {author} {\bibinfo {author} {\bibnamefont {{A. E. van den
  Bogaard}}}\ and\ \bibinfo {author} {\bibfnamefont {E.~E.}\ \bibnamefont
  {Stobberingh}},\ }\href {\doibase 10.1016/S0924-8579(00)00145-X} {\bibfield
  {journal} {\bibinfo  {journal} {Int. J. Antimicrob. Ag.}\ }\textbf {\bibinfo
  {volume} {14}},\ \bibinfo {pages} {327} (\bibinfo {year} {2000})}\BibitemShut
  {NoStop}%
\bibitem [{\citenamefont {{World Health Organization}}(2004)}]{Mal-04}%
  \BibitemOpen
  \bibfield  {author} {\bibinfo {author} {\bibnamefont {{World Health
  Organization}}},\ }\href
  {http://www.who.int/globalchange/publications/en/oeh0401.pdf} {\enquote
  {\bibinfo {title} {Using climate to predict infectious disease outbreaks: A
  review},}\ } (\bibinfo {year} {2004})\BibitemShut {NoStop}%
\bibitem [{\citenamefont {McMichael}\ \emph {et~al.}(2006)\citenamefont
  {McMichael}, \citenamefont {Woodruff},\ and\ \citenamefont {Hales}}]{Mcm-06}%
  \BibitemOpen
  \bibfield  {author} {\bibinfo {author} {\bibfnamefont {A.~J.}\ \bibnamefont
  {McMichael}}, \bibinfo {author} {\bibfnamefont {R.~E.}\ \bibnamefont
  {Woodruff}}, \ and\ \bibinfo {author} {\bibfnamefont {S.}~\bibnamefont
  {Hales}},\ }\href {\doibase 10.1016/S0140-6736(06)68079-3} {\bibfield
  {journal} {\bibinfo  {journal} {The Lancet}\ }\textbf {\bibinfo {volume}
  {367}},\ \bibinfo {pages} {859} (\bibinfo {year} {2006})}\BibitemShut
  {NoStop}%
\bibitem [{\citenamefont {Greger}(2007)}]{Gre-07}%
  \BibitemOpen
  \bibfield  {author} {\bibinfo {author} {\bibfnamefont {M.}~\bibnamefont
  {Greger}},\ }\href {\doibase 10.1080/10408410701647594} {\bibfield  {journal}
  {\bibinfo  {journal} {Crit. Rev. Microbiol.}\ }\textbf {\bibinfo {volume}
  {33}},\ \bibinfo {pages} {243} (\bibinfo {year} {2007})}\BibitemShut
  {NoStop}%
\bibitem [{\citenamefont {Newman}(2002)}]{New-02}%
  \BibitemOpen
  \bibfield  {author} {\bibinfo {author} {\bibfnamefont {M.~E.~J.}\
  \bibnamefont {Newman}},\ }\href@noop {} {\bibfield  {journal} {\bibinfo
  {journal} {Phys. Rev. E}\ }\textbf {\bibinfo {volume} {66}},\ \bibinfo
  {pages} {016128} (\bibinfo {year} {2002})}\BibitemShut {NoStop}%
\bibitem [{\citenamefont {Boccaletti}\ \emph {et~al.}(2006)\citenamefont
  {Boccaletti}, \citenamefont {Latora}, \citenamefont {Moreno}, \citenamefont
  {Chavez},\ and\ \citenamefont {Hwang}}]{Bocc-06}%
  \BibitemOpen
  \bibfield  {author} {\bibinfo {author} {\bibfnamefont {S.}~\bibnamefont
  {Boccaletti}}, \bibinfo {author} {\bibfnamefont {V.}~\bibnamefont {Latora}},
  \bibinfo {author} {\bibfnamefont {Y.}~\bibnamefont {Moreno}}, \bibinfo
  {author} {\bibfnamefont {M.}~\bibnamefont {Chavez}}, \ and\ \bibinfo {author}
  {\bibfnamefont {D.}~\bibnamefont {Hwang}},\ }\href
  {https://doi.org/10.1016/j.physrep.2005.10.009} {\bibfield  {journal}
  {\bibinfo  {journal} {Phys. Rep.}\ }\textbf {\bibinfo {volume} {424}},\
  \bibinfo {pages} {175} (\bibinfo {year} {2006})}\BibitemShut {NoStop}%
\bibitem [{\citenamefont {Newman}(2010)}]{New-10}%
  \BibitemOpen
  \bibfield  {author} {\bibinfo {author} {\bibfnamefont {M.~E.~J.}\
  \bibnamefont {Newman}},\ }\href {\doibase
  10.1093/acprof:oso/9780199206650.001.0001} {\emph {\bibinfo {title}
  {Networks: An Introduction}}}\ (\bibinfo  {publisher} {Oxford University
  Press},\ \bibinfo {year} {2010})\BibitemShut {NoStop}%
\bibitem [{\citenamefont {Castellano}\ and\ \citenamefont
  {Pastor-Satorras}(2010)}]{Cast-10}%
  \BibitemOpen
  \bibfield  {author} {\bibinfo {author} {\bibfnamefont {C.}~\bibnamefont
  {Castellano}}\ and\ \bibinfo {author} {\bibfnamefont {R.}~\bibnamefont
  {Pastor-Satorras}},\ }\href {\doibase 10.1103/PhysRevLett.105.218701}
  {\bibfield  {journal} {\bibinfo  {journal} {Phys. Rev. Lett}\ }\textbf
  {\bibinfo {volume} {105}},\ \bibinfo {pages} {218701} (\bibinfo {year}
  {2010})}\BibitemShut {NoStop}%
\bibitem [{\citenamefont {Pastor-Satorras}\ \emph {et~al.}(2015)\citenamefont
  {Pastor-Satorras}, \citenamefont {Castellano}, \citenamefont {Van~Mieghem},\
  and\ \citenamefont {Vespignani}}]{Pastor-15}%
  \BibitemOpen
  \bibfield  {author} {\bibinfo {author} {\bibfnamefont {R.}~\bibnamefont
  {Pastor-Satorras}}, \bibinfo {author} {\bibfnamefont {C.}~\bibnamefont
  {Castellano}}, \bibinfo {author} {\bibfnamefont {P.}~\bibnamefont
  {Van~Mieghem}}, \ and\ \bibinfo {author} {\bibfnamefont {A.}~\bibnamefont
  {Vespignani}},\ }\href {\doibase https://doi.org/10.1103/RevModPhys.87.925}
  {\bibfield  {journal} {\bibinfo  {journal} {Rev. Mod. Phys.}\ }\textbf
  {\bibinfo {volume} {87}},\ \bibinfo {pages} {925} (\bibinfo {year}
  {2015})}\BibitemShut {NoStop}%
\bibitem [{\citenamefont {Wang}\ \emph {et~al.}(2017)\citenamefont {Wang},
  \citenamefont {Tang}, \citenamefont {Stanley},\ and\ \citenamefont
  {Braunstein}}]{Brau-17}%
  \BibitemOpen
  \bibfield  {author} {\bibinfo {author} {\bibfnamefont {W.}~\bibnamefont
  {Wang}}, \bibinfo {author} {\bibfnamefont {M.}~\bibnamefont {Tang}}, \bibinfo
  {author} {\bibfnamefont {H.~E.}\ \bibnamefont {Stanley}}, \ and\ \bibinfo
  {author} {\bibfnamefont {L.~A.}\ \bibnamefont {Braunstein}},\ }\href@noop {}
  {\bibfield  {journal} {\bibinfo  {journal} {Reports on Progress in Physics}\
  }\textbf {\bibinfo {volume} {80}},\ \bibinfo {pages} {036603} (\bibinfo
  {year} {2017})}\BibitemShut {NoStop}%
\bibitem [{\citenamefont {Grassberger}(1983)}]{Gra-83}%
  \BibitemOpen
  \bibfield  {author} {\bibinfo {author} {\bibfnamefont {P.}~\bibnamefont
  {Grassberger}},\ }\href {\doibase 10.1016/0025-5564(82)90036-0} {\bibfield
  {journal} {\bibinfo  {journal} {Math. Biosci.}\ }\textbf {\bibinfo {volume}
  {63}},\ \bibinfo {pages} {157} (\bibinfo {year} {1983})}\BibitemShut
  {NoStop}%
\bibitem [{\citenamefont {Bailey}(1975)}]{Bai-75}%
  \BibitemOpen
  \bibfield  {author} {\bibinfo {author} {\bibfnamefont {N.~T.~J.}\
  \bibnamefont {Bailey}},\ }\href@noop {} {\emph {\bibinfo {title} {{The
  Mathematical Theory of Infectious Diseases}}}}\ (\bibinfo  {publisher}
  {Griffin, London},\ \bibinfo {year} {1975})\BibitemShut {NoStop}%
\bibitem [{\citenamefont {Miller}(2007)}]{Mil-07}%
  \BibitemOpen
  \bibfield  {author} {\bibinfo {author} {\bibfnamefont {J.~C.}\ \bibnamefont
  {Miller}},\ }\href@noop {} {\bibfield  {journal} {\bibinfo  {journal} {Phys.
  Rev. E}\ }\textbf {\bibinfo {volume} {76}},\ \bibinfo {pages} {010101}
  (\bibinfo {year} {2007})}\BibitemShut {NoStop}%
\bibitem [{\citenamefont {Kenah}\ and\ \citenamefont {Robins}(2007)}]{Ken-07}%
  \BibitemOpen
  \bibfield  {author} {\bibinfo {author} {\bibfnamefont {E.}~\bibnamefont
  {Kenah}}\ and\ \bibinfo {author} {\bibfnamefont {J.~M.}\ \bibnamefont
  {Robins}},\ }\href {\doibase 10.1103/PhysRevE.76.036113} {\bibfield
  {journal} {\bibinfo  {journal} {Phys. Rev. E}\ }\textbf {\bibinfo {volume}
  {76}},\ \bibinfo {pages} {036113} (\bibinfo {year} {2007})}\BibitemShut
  {NoStop}%
\bibitem [{\citenamefont {Lagorio}\ \emph {et~al.}(2009)\citenamefont
  {Lagorio}, \citenamefont {Migueles}, \citenamefont {Braunstein},
  \citenamefont {L\'opez},\ and\ \citenamefont {Macri}}]{Lag-09}%
  \BibitemOpen
  \bibfield  {author} {\bibinfo {author} {\bibfnamefont {C.}~\bibnamefont
  {Lagorio}}, \bibinfo {author} {\bibfnamefont {M.~V.}\ \bibnamefont
  {Migueles}}, \bibinfo {author} {\bibfnamefont {L.~A.}\ \bibnamefont
  {Braunstein}}, \bibinfo {author} {\bibfnamefont {E.}~\bibnamefont {L\'opez}},
  \ and\ \bibinfo {author} {\bibfnamefont {P.}~\bibnamefont {Macri}},\ }\href
  {\doibase 10.1016/j.physa.2008.10.045} {\bibfield  {journal} {\bibinfo
  {journal} {Physica A}\ }\textbf {\bibinfo {volume} {388}},\ \bibinfo {pages}
  {755} (\bibinfo {year} {2009})}\BibitemShut {NoStop}%
\bibitem [{\citenamefont {Meyers}(2007)}]{Mey-07}%
  \BibitemOpen
  \bibfield  {author} {\bibinfo {author} {\bibfnamefont {L.~A.}\ \bibnamefont
  {Meyers}},\ }\href@noop {} {\bibfield  {journal} {\bibinfo  {journal} {Bull.
  Amer. Math. Soc.}\ }\textbf {\bibinfo {volume} {44}},\ \bibinfo {pages} {63}
  (\bibinfo {year} {2007})}\BibitemShut {NoStop}%
\bibitem [{\citenamefont {Ferrari}\ \emph {et~al.}(2006)\citenamefont
  {Ferrari}, \citenamefont {Bansal}, \citenamefont {Meyers},\ and\
  \citenamefont {Bj{\o}rnstad}}]{Fer-06}%
  \BibitemOpen
  \bibfield  {author} {\bibinfo {author} {\bibfnamefont {M.~J.}\ \bibnamefont
  {Ferrari}}, \bibinfo {author} {\bibfnamefont {S.}~\bibnamefont {Bansal}},
  \bibinfo {author} {\bibfnamefont {L.~A.}\ \bibnamefont {Meyers}}, \ and\
  \bibinfo {author} {\bibfnamefont {O.~N.}\ \bibnamefont {Bj{\o}rnstad}},\
  }\href {\doibase 10.1098/rspb.2006.3636} {\bibfield  {journal} {\bibinfo
  {journal} {Proc. R. Soc. London, Ser. B}\ }\textbf {\bibinfo {volume}
  {273}},\ \bibinfo {pages} {2743} (\bibinfo {year} {2006})}\BibitemShut
  {NoStop}%
\bibitem [{\citenamefont {Bansal}\ \emph {et~al.}(2006)\citenamefont {Bansal},
  \citenamefont {Pourbohloul},\ and\ \citenamefont {Meyers}}]{Ban-06}%
  \BibitemOpen
  \bibfield  {author} {\bibinfo {author} {\bibfnamefont {S.}~\bibnamefont
  {Bansal}}, \bibinfo {author} {\bibfnamefont {B.}~\bibnamefont {Pourbohloul}},
  \ and\ \bibinfo {author} {\bibfnamefont {L.~A.}\ \bibnamefont {Meyers}},\
  }\href {\doibase 10.1371/journal.pmed.0030387} {\bibfield  {journal}
  {\bibinfo  {journal} {PLoS Med.}\ }\textbf {\bibinfo {volume} {3}},\ \bibinfo
  {pages} {e387} (\bibinfo {year} {2006})}\BibitemShut {NoStop}%
\bibitem [{\citenamefont {Di~Muro}\ \emph {et~al.}(2018)\citenamefont
  {Di~Muro}, \citenamefont {Alvarez-Zuzek}, \citenamefont {Havlin},\ and\
  \citenamefont {Braunstein}}]{DiMu-18}%
  \BibitemOpen
  \bibfield  {author} {\bibinfo {author} {\bibfnamefont {M.~A.}\ \bibnamefont
  {Di~Muro}}, \bibinfo {author} {\bibfnamefont {L.~G.}\ \bibnamefont
  {Alvarez-Zuzek}}, \bibinfo {author} {\bibfnamefont {S.}~\bibnamefont
  {Havlin}}, \ and\ \bibinfo {author} {\bibfnamefont {L.~A.}\ \bibnamefont
  {Braunstein}},\ }\href@noop {} {\bibfield  {journal} {\bibinfo  {journal}
  {New J. Phys.}\ }\textbf {\bibinfo {volume} {20}},\ \bibinfo {pages} {083025}
  (\bibinfo {year} {2018})}\BibitemShut {NoStop}%
\bibitem [{\citenamefont {Gross}\ \emph {et~al.}(2006)\citenamefont {Gross},
  \citenamefont {Dommar~{D'Lima}},\ and\ \citenamefont {Blasius}}]{Gro-06}%
  \BibitemOpen
  \bibfield  {author} {\bibinfo {author} {\bibfnamefont {T.}~\bibnamefont
  {Gross}}, \bibinfo {author} {\bibfnamefont {C.~J.}\ \bibnamefont
  {Dommar~{D'Lima}}}, \ and\ \bibinfo {author} {\bibfnamefont {B.}~\bibnamefont
  {Blasius}},\ }\href {\doibase 10.1103/PhysRevLett.96.208701} {\bibfield
  {journal} {\bibinfo  {journal} {Phys. Rev. Lett.}\ }\textbf {\bibinfo
  {volume} {96}},\ \bibinfo {pages} {208701} (\bibinfo {year}
  {2006})}\BibitemShut {NoStop}%
\bibitem [{\citenamefont {Lagorio}\ \emph {et~al.}(2011)\citenamefont
  {Lagorio}, \citenamefont {Dickison}, \citenamefont {Vazquez}, \citenamefont
  {Braunstein}, \citenamefont {Macri}, \citenamefont {Migueles}, \citenamefont
  {Havlin},\ and\ \citenamefont {Stanley}}]{Lag-11}%
  \BibitemOpen
  \bibfield  {author} {\bibinfo {author} {\bibfnamefont {C.}~\bibnamefont
  {Lagorio}}, \bibinfo {author} {\bibfnamefont {M.}~\bibnamefont {Dickison}},
  \bibinfo {author} {\bibfnamefont {F.}~\bibnamefont {Vazquez}}, \bibinfo
  {author} {\bibfnamefont {L.~A.}\ \bibnamefont {Braunstein}}, \bibinfo
  {author} {\bibfnamefont {P.~A.}\ \bibnamefont {Macri}}, \bibinfo {author}
  {\bibfnamefont {M.~V.}\ \bibnamefont {Migueles}}, \bibinfo {author}
  {\bibfnamefont {S.}~\bibnamefont {Havlin}}, \ and\ \bibinfo {author}
  {\bibfnamefont {H.~E.}\ \bibnamefont {Stanley}},\ }\href {\doibase
  10.1103/PhysRevE.83.026102} {\bibfield  {journal} {\bibinfo  {journal} {Phys.
  Rev. E}\ }\textbf {\bibinfo {volume} {83}},\ \bibinfo {pages} {026102}
  (\bibinfo {year} {2011})}\BibitemShut {NoStop}%
\bibitem [{\citenamefont {Buono}\ \emph {et~al.}(2012)\citenamefont {Buono},
  \citenamefont {Lagorio}, \citenamefont {Macri},\ and\ \citenamefont
  {Braunstein}}]{Buo-12}%
  \BibitemOpen
  \bibfield  {author} {\bibinfo {author} {\bibfnamefont {C.}~\bibnamefont
  {Buono}}, \bibinfo {author} {\bibfnamefont {C.}~\bibnamefont {Lagorio}},
  \bibinfo {author} {\bibfnamefont {P.~A.}\ \bibnamefont {Macri}}, \ and\
  \bibinfo {author} {\bibfnamefont {L.~A.}\ \bibnamefont {Braunstein}},\ }\href
  {\doibase 10.1016/j.physa.2012.04.002} {\bibfield  {journal} {\bibinfo
  {journal} {Physica A}\ }\textbf {\bibinfo {volume} {391}},\ \bibinfo {pages}
  {4181} (\bibinfo {year} {2012})}\BibitemShut {NoStop}%
\bibitem [{\citenamefont {Valdez}\ \emph {et~al.}(2012)\citenamefont {Valdez},
  \citenamefont {Macri},\ and\ \citenamefont {Braunstein}}]{Val-12}%
  \BibitemOpen
  \bibfield  {author} {\bibinfo {author} {\bibfnamefont {L.~D.}\ \bibnamefont
  {Valdez}}, \bibinfo {author} {\bibfnamefont {P.~A.}\ \bibnamefont {Macri}}, \
  and\ \bibinfo {author} {\bibfnamefont {L.~A.}\ \bibnamefont {Braunstein}},\
  }\href@noop {} {\bibfield  {journal} {\bibinfo  {journal} {Phys. Rev. E}\
  }\textbf {\bibinfo {volume} {85}},\ \bibinfo {pages} {036108} (\bibinfo
  {year} {2012})}\BibitemShut {NoStop}%
\bibitem [{\citenamefont {Buono}\ \emph {et~al.}(2013)\citenamefont {Buono},
  \citenamefont {Vazquez}, \citenamefont {Macri},\ and\ \citenamefont
  {Braunstein}}]{Buo-13}%
  \BibitemOpen
  \bibfield  {author} {\bibinfo {author} {\bibfnamefont {C.}~\bibnamefont
  {Buono}}, \bibinfo {author} {\bibfnamefont {F.}~\bibnamefont {Vazquez}},
  \bibinfo {author} {\bibfnamefont {P.~A.}\ \bibnamefont {Macri}}, \ and\
  \bibinfo {author} {\bibfnamefont {L.~A.}\ \bibnamefont {Braunstein}},\
  }\href@noop {} {\bibfield  {journal} {\bibinfo  {journal} {Phys. Rev. E}\
  }\textbf {\bibinfo {volume} {88}},\ \bibinfo {pages} {022813} (\bibinfo
  {year} {2013})}\BibitemShut {NoStop}%
\bibitem [{\citenamefont {Eastwood}\ \emph {et~al.}(2010)\citenamefont
  {Eastwood}, \citenamefont {Durrheim}, \citenamefont {Butler},\ and\
  \citenamefont {Jon}}]{Eas-10}%
  \BibitemOpen
  \bibfield  {author} {\bibinfo {author} {\bibfnamefont {K.}~\bibnamefont
  {Eastwood}}, \bibinfo {author} {\bibfnamefont {D.~N.}\ \bibnamefont
  {Durrheim}}, \bibinfo {author} {\bibfnamefont {M.}~\bibnamefont {Butler}}, \
  and\ \bibinfo {author} {\bibfnamefont {E.~A.}\ \bibnamefont {Jon}},\ }\href
  {\doibase 10.3201/eid1608.100132} {\bibfield  {journal} {\bibinfo  {journal}
  {Emerg. Infect. Dis.}\ }\textbf {\bibinfo {volume} {16}},\ \bibinfo {pages}
  {1211} (\bibinfo {year} {2010})}\BibitemShut {NoStop}%
\bibitem [{\citenamefont {Cattuto}\ \emph {et~al.}(2010)\citenamefont
  {Cattuto}, \citenamefont {den Broeck}, \citenamefont {Barrat}, \citenamefont
  {Colizza}, \citenamefont {Pinton},\ and\ \citenamefont
  {Vespignani}}]{Catt-10}%
  \BibitemOpen
  \bibfield  {author} {\bibinfo {author} {\bibfnamefont {C.}~\bibnamefont
  {Cattuto}}, \bibinfo {author} {\bibfnamefont {W.~V.}\ \bibnamefont {den
  Broeck}}, \bibinfo {author} {\bibfnamefont {A.}~\bibnamefont {Barrat}},
  \bibinfo {author} {\bibfnamefont {V.}~\bibnamefont {Colizza}}, \bibinfo
  {author} {\bibfnamefont {J.~F.}\ \bibnamefont {Pinton}}, \ and\ \bibinfo
  {author} {\bibfnamefont {A.}~\bibnamefont {Vespignani}},\ }\href {\doibase
  10.1371/journal.pone.0011596} {\bibfield  {journal} {\bibinfo  {journal}
  {PLoS ONE}\ }\textbf {\bibinfo {volume} {5}},\ \bibinfo {pages} {e11596}
  (\bibinfo {year} {2010})}\BibitemShut {NoStop}%
\bibitem [{\citenamefont {Karsai}\ \emph {et~al.}(2011)\citenamefont {Karsai},
  \citenamefont {Kivel{\"a}}, \citenamefont {Pan}, \citenamefont {Kaski},
  \citenamefont {Kert{\'e}sz}, \citenamefont {Barab{\'a}si},\ and\
  \citenamefont {Saram{\"a}ki}}]{Kars-11}%
  \BibitemOpen
  \bibfield  {author} {\bibinfo {author} {\bibfnamefont {M.}~\bibnamefont
  {Karsai}}, \bibinfo {author} {\bibfnamefont {M.}~\bibnamefont {Kivel{\"a}}},
  \bibinfo {author} {\bibfnamefont {R.~K.}\ \bibnamefont {Pan}}, \bibinfo
  {author} {\bibfnamefont {K.}~\bibnamefont {Kaski}}, \bibinfo {author}
  {\bibfnamefont {J.}~\bibnamefont {Kert{\'e}sz}}, \bibinfo {author}
  {\bibfnamefont {A.~L.}\ \bibnamefont {Barab{\'a}si}}, \ and\ \bibinfo
  {author} {\bibfnamefont {J.}~\bibnamefont {Saram{\"a}ki}},\ }\href@noop {}
  {\bibfield  {journal} {\bibinfo  {journal} {Phys. Rev. E}\ }\textbf {\bibinfo
  {volume} {83}},\ \bibinfo {pages} {025102} (\bibinfo {year}
  {2011})}\BibitemShut {NoStop}%
\bibitem [{\citenamefont {Stehle\'e}\ \emph {et~al.}(2011)\citenamefont
  {Stehle\'e}, \citenamefont {Barrat}, \citenamefont {Cattuto}, \citenamefont
  {Pinton}, \citenamefont {Isella},\ and\ \citenamefont {den
  Broeck}}]{Steh-11}%
  \BibitemOpen
  \bibfield  {author} {\bibinfo {author} {\bibfnamefont {J.}~\bibnamefont
  {Stehle\'e}}, \bibinfo {author} {\bibfnamefont {A.}~\bibnamefont {Barrat}},
  \bibinfo {author} {\bibfnamefont {C.}~\bibnamefont {Cattuto}}, \bibinfo
  {author} {\bibfnamefont {J.~F.}\ \bibnamefont {Pinton}}, \bibinfo {author}
  {\bibfnamefont {L.}~\bibnamefont {Isella}}, \ and\ \bibinfo {author}
  {\bibfnamefont {W.~V.}\ \bibnamefont {den Broeck}},\ }\href@noop {}
  {\bibfield  {journal} {\bibinfo  {journal} {J. Theor. Biol.}\ }\textbf
  {\bibinfo {volume} {271}},\ \bibinfo {pages} {166} (\bibinfo {year}
  {2011})}\BibitemShut {NoStop}%
\bibitem [{\citenamefont {Valdez}\ \emph {et~al.}(2013)\citenamefont {Valdez},
  \citenamefont {Buono}, \citenamefont {Macri},\ and\ \citenamefont
  {Braunstein}}]{Val-13}%
  \BibitemOpen
  \bibfield  {author} {\bibinfo {author} {\bibfnamefont {L.~D.}\ \bibnamefont
  {Valdez}}, \bibinfo {author} {\bibfnamefont {C.}~\bibnamefont {Buono}},
  \bibinfo {author} {\bibfnamefont {P.~A.}\ \bibnamefont {Macri}}, \ and\
  \bibinfo {author} {\bibfnamefont {L.~A.}\ \bibnamefont {Braunstein}},\
  }\href@noop {} {\bibfield  {journal} {\bibinfo  {journal} {FRACTALS}\
  }\textbf {\bibinfo {volume} {21}},\ \bibinfo {pages} {1350019} (\bibinfo
  {year} {2013})}\BibitemShut {NoStop}%
\bibitem [{\citenamefont {Molloy}\ and\ \citenamefont {Reed}(1995)}]{Moll-95}%
  \BibitemOpen
  \bibfield  {author} {\bibinfo {author} {\bibfnamefont {M.}~\bibnamefont
  {Molloy}}\ and\ \bibinfo {author} {\bibfnamefont {B.}~\bibnamefont {Reed}},\
  }\href {\doibase 10.1002/rsa.3240060204} {\bibfield  {journal} {\bibinfo
  {journal} {Random Struct. Algor.}\ }\textbf {\bibinfo {volume} {6}},\
  \bibinfo {pages} {161} (\bibinfo {year} {1995})}\BibitemShut {NoStop}%
\bibitem [{\citenamefont {Braunstein}\ \emph {et~al.}(2007)\citenamefont
  {Braunstein}, \citenamefont {Wu}, \citenamefont {Chen}, \citenamefont
  {Buldyrev}, \citenamefont {Kalisy}, \citenamefont {Sreenivasan},
  \citenamefont {Cohen}, \citenamefont {López}, \citenamefont {Havlin},\ and\
  \citenamefont {Stanley}}]{Brau-07}%
  \BibitemOpen
  \bibfield  {author} {\bibinfo {author} {\bibfnamefont {L.~A.}\ \bibnamefont
  {Braunstein}}, \bibinfo {author} {\bibfnamefont {Z.}~\bibnamefont {Wu}},
  \bibinfo {author} {\bibfnamefont {Y.}~\bibnamefont {Chen}}, \bibinfo {author}
  {\bibfnamefont {S.~V.}\ \bibnamefont {Buldyrev}}, \bibinfo {author}
  {\bibfnamefont {T.}~\bibnamefont {Kalisy}}, \bibinfo {author} {\bibfnamefont
  {S.}~\bibnamefont {Sreenivasan}}, \bibinfo {author} {\bibfnamefont
  {R.}~\bibnamefont {Cohen}}, \bibinfo {author} {\bibfnamefont
  {E.}~\bibnamefont {López}}, \bibinfo {author} {\bibfnamefont
  {S.}~\bibnamefont {Havlin}}, \ and\ \bibinfo {author} {\bibfnamefont {H.~E.}\
  \bibnamefont {Stanley}},\ }\href@noop {} {\bibfield  {journal} {\bibinfo
  {journal} {International Journal of Bifurcation and Chaos}\ }\textbf
  {\bibinfo {volume} {17}},\ \bibinfo {pages} {2215} (\bibinfo {year}
  {2007})}\BibitemShut {NoStop}%
\bibitem [{\citenamefont {Newman}\ \emph {et~al.}(2001)\citenamefont {Newman},
  \citenamefont {Strogatz},\ and\ \citenamefont {Watts}}]{New-01}%
  \BibitemOpen
  \bibfield  {author} {\bibinfo {author} {\bibfnamefont {M.~E.~J.}\
  \bibnamefont {Newman}}, \bibinfo {author} {\bibfnamefont {S.~H.}\
  \bibnamefont {Strogatz}}, \ and\ \bibinfo {author} {\bibfnamefont {D.~J.}\
  \bibnamefont {Watts}},\ }\href@noop {} {\bibfield  {journal} {\bibinfo
  {journal} {Phys. Rev. E}\ }\textbf {\bibinfo {volume} {64}},\ \bibinfo
  {pages} {026118} (\bibinfo {year} {2001})}\BibitemShut {NoStop}%
\bibitem [{\citenamefont {Newman}(2003)}]{New-03}%
  \BibitemOpen
  \bibfield  {author} {\bibinfo {author} {\bibfnamefont {M.~E.~J.}\
  \bibnamefont {Newman}},\ }\href {\doibase 10.1137/S003614450342480}
  {\bibfield  {journal} {\bibinfo  {journal} {SIAM Rev.}\ }\textbf {\bibinfo
  {volume} {45}},\ \bibinfo {pages} {167} (\bibinfo {year} {2003})}\BibitemShut
  {NoStop}%
\end{thebibliography}
\end{document}